\documentclass[twocolumn,english,superscriptaddress]{revtex4}
\usepackage[T1]{fontenc}
\usepackage[latin9]{inputenc}
\usepackage{amsmath}
\usepackage{amssymb}
\usepackage{stmaryrd}
\usepackage{graphicx}
\usepackage{esint}
\usepackage{xcolor}

\makeatletter

\newcommand*\LyXThinSpace{\,\hspace{0pt}}

\@ifundefined{textcolor}{}
{%
 \definecolor{BLACK}{gray}{0}
 \definecolor{WHITE}{gray}{1}
 \definecolor{RED}{rgb}{1,0,0}
 \definecolor{GREEN}{rgb}{0,1,0}
 \definecolor{BLUE}{rgb}{0,0,1}
 \definecolor{CYAN}{cmyk}{1,0,0,0}
 \definecolor{MAGENTA}{cmyk}{0,1,0,0}
 \definecolor{YELLOW}{cmyk}{0,0,1,0}
}

\usepackage{babel}

\makeatother

\usepackage{babel}
\begin{document}

\title{Impact of disorder on the superconducting transition temperature
near a Lifshitz transition}

\author{Thaís V. Trevisan}

\affiliation{School of Physics and Astronomy, University of Minnesota, Minneapolis
55455, USA}

\affiliation{Instituto de Física Gleb Wataghin, Unicamp, Rua Sérgio Buarque de
Holanda, 777, CEP 13083-859 Campinas, SP, Brazil}

\author{Michael Schütt}

\affiliation{School of Physics and Astronomy, University of Minnesota, Minneapolis
55455, USA}

\author{Rafael M. Fernandes}

\affiliation{School of Physics and Astronomy, University of Minnesota, Minneapolis
55455, USA}
\begin{abstract}
Multi-band superconductivity is realized in a plethora of systems,
from high-temperature superconductors to very diluted superconductors.
While several properties of multi-band superconductors can be understood
as straightforward generalizations of their single-band counterparts,
recent works have unveiled rather unusual behaviors unique to the
former case. In this regard, a regime that has received significant
attention is that near a Lifshitz transition, in which one of the
bands crosses the Fermi level. In this work, we investigate how impurity
scattering $\tau^{-1}$ affects the superconducting transition temperature
$T_{c}$ across a Lifshitz transition, in the regime where intra-band
pairing is dominant and inter-band pairing is subleading. This is
accomplished by deriving analytic asymptotic expressions for $T_{c}$
and $\partial T_{c}/\partial\tau^{-1}$ in a two-dimensional two-band
system. When the inter-band pairing interaction is repulsive, we find
that, despite the incipient nature of the band crossing the Fermi
level, inter-band impurity scattering is extremely effective in breaking
Cooper pairs, making $\partial T_{c}/\partial\tau^{-1}$ quickly approach
the limiting Abrikosov-Gor'kov value of the high-density regime. In
contrast, when the inter-band pairing interaction is attractive, pair-breaking
is much less efficient, affecting $T_{c}$ only mildly at the vicinity
of the Lifshitz transition. The consequence of this general result
is that the behavior of $T_{c}$ across a Lifshitz transition can
be qualitatively changed in the presence of strong enough disorder:
instead of displaying a sharp increase across the Lifshitz transition,
as in the clean case, $T_{c}$ can actually display a maximum and
be suppressed at the Lifshitz transition. These results shed new light
on the non-trivial role of impurity scattering in multi-band superconductors.
\end{abstract}
\maketitle

\section{Introduction}

Just a few years after the development of the BCS theory of superconductivity,
an extension of this model to multi-band superconductors (SC) was
proposed by Suhl \emph{et al. }\cite{Suhl} and Moskalenko \cite{Moskalenko}
to investigate the consequences of overlapping bands in the superconducting
state of certain transition metals. Indeed, multi-band superconductivity
should be common among materials in which multiple electronic $d$
orbitals are occupied, and whose crystal field splittings are not
too large. Currently, there are many known multi-band superconductors,
ranging from conventional superconductors such as MgB$_{2}$ \cite{MgB2},
NbSn$_{3}$ \cite{Nb3Sn}, and NbSe$_{2}$ \cite{NbSe2}, to unconventional
superconductors such as BaFe$_{2}$As$_{2}$ \cite{pnictides}, Sr$_{2}$RuO$_{4}$
\cite{ruthenates}, and CeCoIn$_{5}$ \cite{heavy_fermion}. More
recently, multi-band superconductivity has been demonstrated in bulk
SrTiO$_{3}$ \cite{Binnig80,Behnia_fermi_surfaces} and in LaAlO$_{3}$/SrTiO$_{3}$
heterostructures \cite{Ilani12}, although the microscopic origin
of superconductivity in these systems remains hotly debated \cite{Gorkov16,Balatsky_QCP,Lonzarich14,PLee16,Verri18}.
Theoretically, several recent studies have unveiled unique properties
of multi-band superconductors that are not realized in their single-band
counterparts \cite{Geyer10,Efremov11,Komendova12,Babaev05,Maiti13,Schmalian15,Komendova15,Babaev11}.

An interesting regime in multi-band superconductors is when one of
the bands is incipient, i.e. its bottom (or top) is just below (or
above) the Fermi level. The appearance or disappearance of a Fermi
pocket from the Fermi surface is often called a Lifshitz transition
(LT) \cite{Lifshitz}. Note that, in its original conception, a LT
referred to a change in the topology of the Fermi surface from open
to closed. However, given the widespread use of this term to denote
also the situation of a band crossing the Fermi level, we will here
use LT to refer to the latter case. Near a LT, the energy scale of
the pairing interaction is larger than the Fermi energy of the incipient
band, which may lead to interesting new behaviors \cite{Bianconi94a,Fernandes13,Bianconi14,Chubukov16,Balatsky_2bands,Hirschfeld16,Bang14,Hirschfeld15,Bianconi10,Koshelev17,Valentinis16}.

Experimentally, tuning a multi-band superconductor to a LT has been
achieved by doping, gating, and even pressure. For instance, such
a LT has been shown to take place in the phase diagrams of Ba(Fe$_{1-x}$Co$_{x}$)$_{2}$As$_{2}$
\cite{Liu11}, pressurized KFe$_{2}$As$_{2}$ \cite{kfe2as2}, SrTiO$_{3-\delta}$
\cite{Behnia_fermi_surfaces}, and gated SrTiO$_{3}$/LaAlO$_{3}$
\cite{Ilani12}. Theoretically, the goal is to relate the thermodynamic
properties of the SC across the LT transition with the microscopic
properties of the gap function, in order to shed light on the mechanisms
involved in the pairing problem. Take, for instance, the case of Ba(Fe$_{1-x}$Co$_{x}$)$_{2}$As$_{2}$:
the superconducting transition temperature $T_{c}$ was found to vanish
when the hole pockets sank below the Fermi level, indicating the dominance
of inter-band pairing over intra-band pairing \cite{Liu11}. The latter
would be expected to dominate if the standard electron-phonon interaction
was the pairing glue. The situation, however, is much less clear in
SrTiO$_{3-\delta}$ and gated LaAlO$_{3}$/SrTiO$_{3}$ \cite{Behnia_fermi_surfaces,Ilani12}:
there, superconductivity is quite well established in the single-band
regime, indicating dominant intra-band pairing. However, $T_{c}$
is actually suppressed across the LT, once the second band crosses
the Fermi level. Such a behavior is at odds with general theoretical
expectations that $T_{c}$ should increase across a LT since the extra
band provides more carriers to be part of the SC state \cite{Bianconi94a,Fernandes13,Valentinis16}.

In this paper, we investigate how disorder affects $T_{c}$ and the
gap functions across a two-band LT. We argue that the impact of disorder
is fundamentally different depending on whether the inter-band pairing
interaction is repulsive or attractive. In the former case, inter-band
impurity scattering is strongly pair-breaking, implying that once
the second band becomes part of the Fermi surface, pair-breaking effects
become more substantial. Interestingly, crossing the LT leads to a
change in the pairing symmetry from sign-changing gaps between the
two bands to same-sign gaps \cite{Trevisan_Schutt_Fernandes}. These
effects, in contrast, do not happen for an attractive inter-band interaction.

In our previous work \cite{Trevisan_Schutt_Fernandes}, this problem
was solved numerically in 3D and in 2D in the dirty limit, and applied
to the particular cases of SrTiO$_{3-\delta}$ and gated LaAlO$_{3}$/SrTiO$_{3}$.
Here, we instead focus on general analytical asymptotic results for
small impurity scattering in 2D, which leads to important insights
on the mechanisms involved. We obtain not only analytic expressions
for $T_{c}$, but also for the rate of change of $T_{c}$ with respect
to inter-band impurity scattering $\tau_{\mathrm{inter}}^{-1}$, $\partial T_{c}/\partial\tau_{\mathrm{inter}}^{-1}$.
The latter is derived by using a technique based on Hellmann-Feynman
theorem, following the seminal work of Ref. \cite{Rainer73}. Starting
in the high-density regime, where the system has long crossed the
Lifshitz transition, we recover the well-known result for identical
bands that $\partial T_{c}/\partial\tau_{\mathrm{inter}}^{-1}=0$
for attractive inter-band pairing (sign-preserving $s^{++}$ superconducting
state), and $\partial T_{c}/\partial\tau_{\mathrm{inter}}^{-1}=-\pi/4$
(the universal Abrikosov-Gor'kov value) for repulsive inter-band scattering
(sign-changing $s^{+-}$ superconducting state). Deviations from this
fine-tuned condition of identical bands with identical intra-band
pairing interactions leads to a reduction of the $T_{c}$ suppression
in the $s^{+-}$ case, and an enhancement of the $T_{c}$ suppression
in the $s^{++}$ case. When the system is well inside the single-band
regime, i.e. well before crossing the Lifshitz transition, the suppression
rate $\partial T_{c}/\partial\tau_{\mathrm{inter}}^{-1}$ is very
small regardless of the sign of the inter-band pairing. The interesting
behavior takes place in the vicinity of the Lifshitz transition. For
the $s^{+-}$ state, we show that $\partial T_{c}/\partial\tau_{\mathrm{inter}}^{-1}$
is strongly suppressed and quickly approaches the high-density value,
even in the regime where the second band is only incipient. This contrasts
to the behavior of the $s^{++}$ state, in which $\partial T_{c}/\partial\tau_{\mathrm{inter}}^{-1}$
has a small minimum at the Lifshitz transition, before it increases
towards the high-density value. 

The paper is organized in the following way: to introduce the model,
we start in Sec.\ref{sec_clean} with a clean two-band superconductor,
solving the pairing problem both numerically and analytically. In
Sec.\ref{sec_dirty}, we generalize the model to include non-magnetic
random impurities. Sec. \ref{sec_analytics} presents the analytic
asymptotic solutions of the dirty superconductor across a LT both
in the high-density regime and in the dilute regime. In Sec.\ref{sec_concl}
we summarize our conclusions. Appendices \ref{Appendix1}, \ref{Appendix2} and \ref{Appendix3}
provide more details about the analytic calculations performed in
the main text.

\section{Clean two-band superconductor \label{sec_clean}}

\subsection{Gap equations}

\begin{figure}[b!]
\centering \includegraphics[width=0.65\columnwidth]{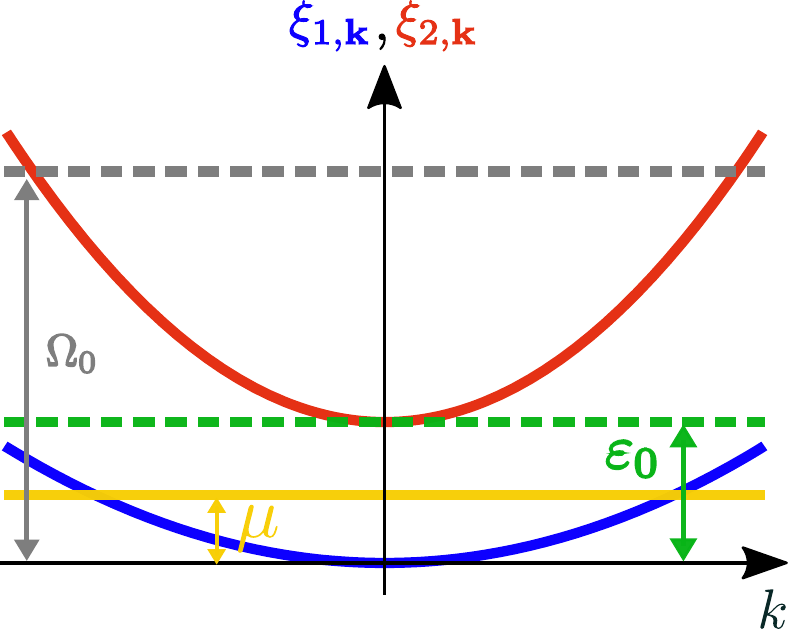}
\caption{(color online) Illustration of the two-band model used in this work. Two electron-like
parabolic and concentric bands are displaced by an energy $\varepsilon_{0}>0$.
Their occupations are controlled by the chemical potential $\mu>0$.
When $\mu$ becomes larger than $\varepsilon_{0}$, the second band
becomes populated, signaling a Lifshitz transition (LT). }
\label{s1fig1} 
\end{figure}

The two-band superconducting system that we study here is described
by the Hamiltonian:
\begin{align}
H_{0} & =\sum\limits _{\mathbf{k},i,\sigma}\xi_{i,\mathbf{k}}^{\null}\,c_{i,\mathbf{k}\sigma}^{\dag}c_{i,\mathbf{k}\sigma}^{\null}\nonumber \\
 & +\sum\limits _{\mathbf{k},\mathbf{k'},i,j}V_{ij}c_{i,\mathbf{k}\uparrow}^{\dag}c_{i,-\mathbf{k}\downarrow}^{\dag}c_{j,-\mathbf{k'}\downarrow}^{\null}c_{j,\mathbf{k'}\uparrow}^{\null}\text{ ,}\label{s1eq1-1}
\end{align}
where $c_{j,\mathbf{k}\sigma}^{\dag}$ and $c_{j,\mathbf{k}\sigma}^{\null}$
are the operators that create and annihilate, respectively, an electron
in band $i$ ($i=1,2$), with momentum $\mathbf{k}$ and spin $\sigma$.
As in Refs. \cite{Fernandes13,Trevisan_Schutt_Fernandes}, we consider
parabolic electron-like bands $\xi_{1,\mathbf{k}}=\frac{k^{2}}{2m_{1}}-\mu$
and $\xi_{2,\mathbf{k}}=\frac{k^{2}}{2m_{2}}-\mu+\varepsilon_{0}$,
as illustrated in Fig. \ref{s1fig1}. The bottom of band $1$, $W_{1}=-\mu$,
is split from the bottom of band $2$, $W_{2}=-\mu+\varepsilon_{0}$,
by the energy scale $\varepsilon_{0}>0$. The chemical potential $\mu>0$
is a control parameter in our model, which tunes the system through
a Lifshitz transition (LT) at $\mu=\varepsilon_{0}$.

The pairing interaction is described by the matrix $V_{ij}$ and contains
both (momentum-independent) intra-band pairing, $V_{11}$ and $V_{22}$
which do not need to be necessarily equal, and inter-band pairing,
$V_{12}=V_{21}$. As a result, the isotropic SC gap $\Delta_{i}$
in band $i$ is given by:

\noindent 
\begin{equation}
\Delta_{i}=-\sum\limits _{\mathbf{k},j}V_{ij}\left\langle c_{j,-\mathbf{k}\downarrow}c_{j,\mathbf{k}\uparrow}\right\rangle \text{ .}\label{s1eq3}
\end{equation}
yielding the usual mean-field Hamiltonian: 
\begin{align}
H_{0}= & \sum\limits _{\mathbf{k},i,\sigma}\xi_{i,\mathbf{k}}\,c_{i,\mathbf{k}\sigma}^{\dag}c_{i,\mathbf{k}\sigma}^{\null}\nonumber \\
 & -\sum\limits _{\mathbf{k},i}\left(\Delta_{i}\,c_{i,\mathbf{k}\uparrow}^{\dag}c_{i,-\mathbf{k}\downarrow}^{\dag}+h.c.\right)\text{ ,}\label{s1eq2}
\end{align}

Before introducing disorder, we rederive the results for $T_{c}$
of a clean two-band system across a LT (see also Ref. \cite{Fernandes13}
and references therein). Introducing the Nambu spinor $\hat{\psi}_{\mathbf{k}}^{\dagger}=\left(c_{1,\mathbf{k}\uparrow}^{\dag}\,c_{1,-\mathbf{k}\downarrow}^{\null}\,c_{2,\mathbf{k}\uparrow}^{\dag}\,c_{2,-\mathbf{k}\downarrow}^{\null}\right)$,
we can readily obtain the normal and anomalous Green's functions of
band $i$, $\mathcal{G}_{i}$ and $\mathcal{F}_{i}$, which appear
in the Nambu's Green's function $\hat{\mathcal{G}}_{0}$ as:

\noindent 
\begin{equation}
\hat{\mathcal{G}}_{0}(\mathbf{k},\omega_{n})=\begin{pmatrix}\mathcal{G}_{1,0} & \mathcal{F}_{1,0} & 0 & 0\\
\mathcal{F}_{1,0} & -\mathcal{G}_{1,0}^{*} & 0 & 0\\
0 & 0 & \mathcal{G}_{2,0} & \mathcal{F}_{2,0}\\
0 & 0 & \mathcal{F}_{2,0} & -\mathcal{G}_{2,0}^{*}
\end{pmatrix}\text{ ,}\label{s1eq5}
\end{equation}

We find:

\noindent 
\begin{equation}
\mathcal{G}_{i,0}(\mathbf{k},\omega_{n})=-\frac{i\omega_{n}+\xi_{i,\mathbf{k}}}{\omega_{n}^{2}+\xi_{i,\mathbf{k}}^{2}+\Delta_{i}^{2}}\text{ ,}\label{s1eq6}
\end{equation}

\noindent and 
\begin{equation}
\mathcal{F}_{i,0}(\mathbf{k},\omega_{n})=\frac{\Delta_{i}}{\omega_{n}^{2}+\xi_{i,\mathbf{k}}^{2}+\Delta_{i}^{2}}\text{ .}\label{s1eq7}
\end{equation}

The latter is related to the pair expectation value, $\left\langle c_{i,-\mathbf{k}\downarrow}c_{i,\mathbf{k}\uparrow}\right\rangle =T\sum_{n}\mathcal{F}_{i,0}(\mathbf{k},\omega_{n})$,
from which we can derive the gap equation:
\begin{equation}
\Delta_{i}=\pi T\sum\limits _{j,n}\lambda_{ij}\Delta_{j}\left\langle \frac{1}{\omega_{n}^{2}+\xi^{2}+\Delta_{j}^{2}}\right\rangle _{j}^{\Omega_{0}}\!\!\text{ .}\label{s1eq9}
\end{equation}

\noindent Here, we introduced the dimensionless coupling constants
$\lambda_{ij}=-\rho_{j,0}V_{ij}$, such that positive and negative
$\lambda_{ij}$ correspond to attraction and repulsion, respectively.
We also defined the notation:

\begin{equation}
\left\langle \mathcal{O}(\xi)\right\rangle _{i}^{\xi_{c}}\equiv\frac{1}{\pi\rho_{i,0}}\int\limits _{W_{i}}^{\xi_{c}}d\xi\rho_{i}(\xi)\mathcal{O}(\xi)\text{ ,}\label{integral}
\end{equation}

\noindent where $\mathcal{O}(\xi)$ is an arbitrary function of energy,
$\xi_{c}$ denotes the upper cutoff of the integral, and $W_{i}$
denotes the bottom of band $i$. In the gap equation, the upper limit
of the integration corresponds to the energy cutoff of the pairing
interaction, $\Omega_{0}$, which plays a similar role as the Debye
frequency in the standard BCS approach. Finally, $\rho_{i}(\xi)$
is the density of states \textit{\emph{per spin}} of band $i$, and
$\rho_{i,0}\equiv\rho_{i}(W_{i}+\varepsilon_{0})$. Since we have
parabolic bands, $\rho_{i}(\xi)=\frac{m_{i}}{2\pi}$ for the 2D case
and $\rho_{i}(\xi)=\frac{(2m_{i})^{3/2}\sqrt{\varepsilon_{0}}}{4\pi^{2}}\sqrt{\frac{\xi-W_{i}}{\varepsilon_{0}}}$
for the 3D case, yielding $\rho_{i,0}=\frac{m_{i}}{2\pi}$ and $\rho_{i,0}=\frac{(2m_{i})^{3/2}\sqrt{\varepsilon_{0}}}{4\pi^{2}}$,
respectively. The linearized gap equation follows directly from Eq.
(\ref{s1eq9}):

\noindent 
\begin{equation}
\begin{pmatrix}\Delta_{1}\\
\Delta_{2}
\end{pmatrix}=\begin{pmatrix}\lambda_{11} & \lambda_{12}\\
\lambda_{21} & \lambda_{22}
\end{pmatrix}\hat{A}_{\mathrm{clean}}(\mu,T_{c})\begin{pmatrix}\Delta_{1}\\
\Delta_{2}
\end{pmatrix}\text{ .}\label{s1eq10}
\end{equation}

\begin{figure*}[t!]
\centering \includegraphics[width=0.9\linewidth]{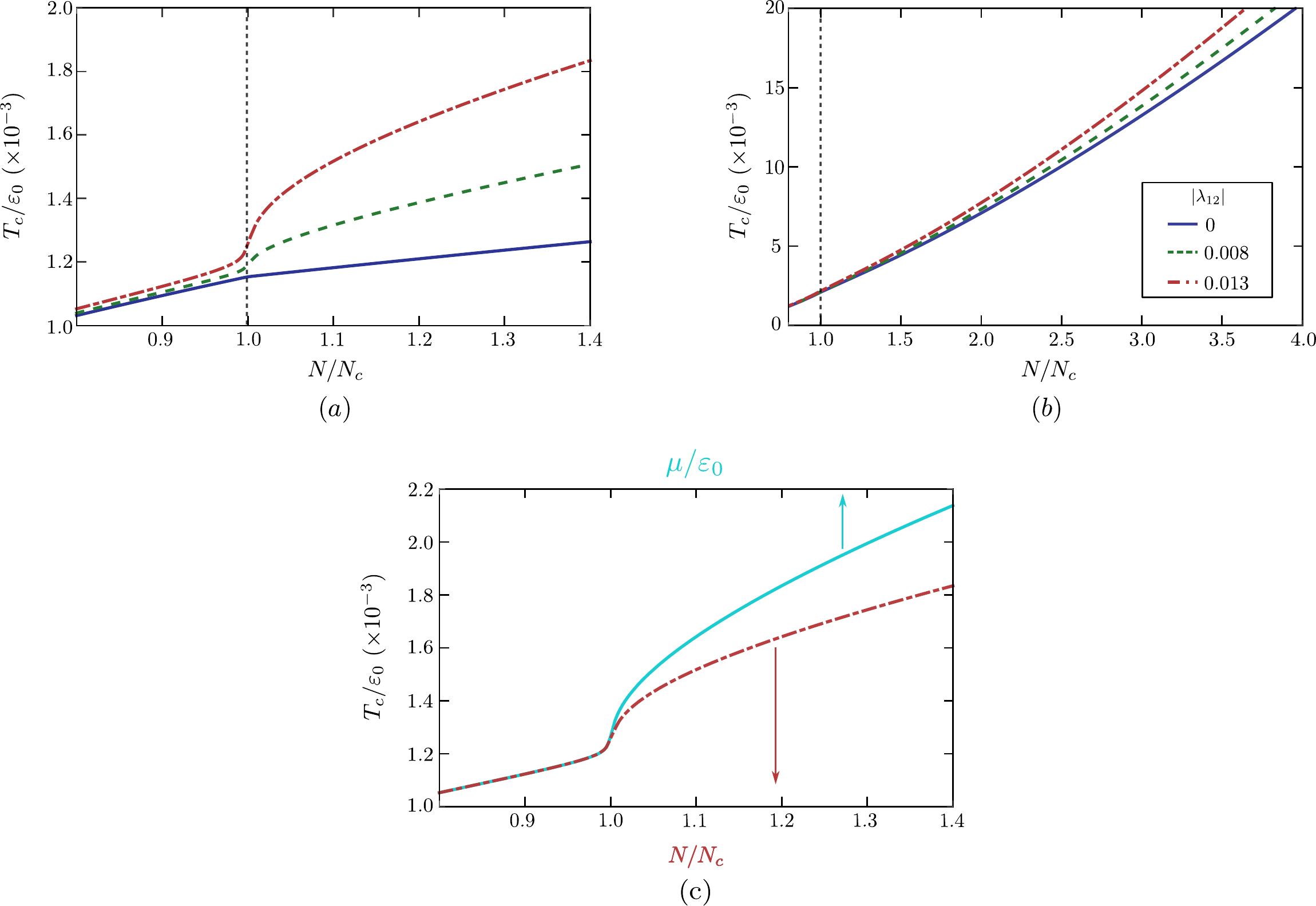}
\caption{(color online) Phase diagram of the clean two-band SC: the first two panels show
$T_{c}$ as function of the occupation number $N$ for (a) 2D bands
and (b) 3D bands, for several values of the parameter $\lambda_{12}$.
In panel (c), we compare $T_{c}$ as a function of $N$ with $T_{c}$
as a function of the chemical potential $\mu(T_{c})$ for the 2D case
with $|\lambda_{12}|=0.013$. In all panels, $\lambda_{11}=\lambda_{22}=0.13$
and $\rho_{2,0}=\rho_{1,0}$. Note that $T_{c}$ is normalized by
the energy displacement between the bands $\varepsilon_{0}$ and $N$
is normalized by the critical occupation number $N_{c}$ at which
the LT takes place.}
\label{clean} 
\end{figure*}

\noindent where $\hat{A}_{\mathrm{clean}}$ has matrix elements:
\begin{align}
\left(\hat{A}_{\mathrm{clean}}\right)_{ij} & =\delta_{ij}\pi T_{c}\sum\limits _{n}\left\langle \frac{1}{\omega_{n}^{2}+\xi^{2}}\right\rangle _{i}^{\Omega_{0}}\nonumber \\
 & =\delta_{ij}\,\frac{\pi}{2}\left\langle \frac{1}{\xi}\,\text{tanh}\left(\frac{\xi}{2T_{c}}\right)\right\rangle _{i}^{\Omega_{0}}\text{ .}\label{s1eq11}
\end{align}

Equation (\ref{s1eq10}) defines an eigenvalue problem. $T_{c}$,
as a function of the chemical potential $\mu$, is determined when
the largest eigenvalue of $\hat{\lambda}\hat{A}_{\mathrm{clean}}$
equals $1$, where $(\hat{\lambda})_{ij}=\lambda_{ij}$. This is given
by: 

\begin{equation}
\prod_{i=1,2}\left[\left(\hat{A}_{\mathrm{clean}}\right)_{ii}\det\left(\hat{\lambda}\right)-\lambda_{\bar{i}\bar{i}}\right]=\lambda_{12}\lambda_{21}\text{ ,}\label{gap_eq}
\end{equation}

\noindent as long as $\det\left(\hat{\lambda}\right)=\lambda_{11}\lambda_{22}-\lambda_{12}\lambda_{21}\neq0$.
Here, we introduced the notation $\bar{i}=1(2)$ for $i=2(1)$. It
is clear that the equations depend only on the product $\lambda_{12}\lambda_{21}$,
i.e. only on the square of the inter-band interaction $V_{12}^{2}$.
As a result, $T_{c}(\mu)$ is independent of whether the inter-band
pairing interaction is repulsive or attractive \cite{Fernandes13,Valentinis16}.
On the other hand, the sign of the off-diagonal term $\lambda_{12}$
(which by definition is the same as the sign of $\lambda_{21}$ and
the opposite of $V_{12}$) determines the eigenvector corresponding
to the largest eigenvalue of $\hat{\lambda}\hat{A}$ . When $\lambda_{12}>0$,
this eigenvector is such that $\Delta_{1}$ and $\Delta_{2}$ have
the same sign, corresponding to a \textit{\emph{conventional}}\emph{
$s^{++}$ }\textit{\emph{SC state}}. When $\lambda_{12}<0$, $\Delta_{1}$
and $\Delta_{2}$ acquire opposite signs, corresponding to an \textit{\emph{unconventional}}\emph{
$s^{+-}$ }\textit{\emph{SC state}}

It is important to note that the chemical potential $\mu$ that appears
in the gap equation is not the $T=0$ chemical potential, but actually
$\mu(T_{c})$. Close to the LT, because the Fermi energy is small,
$\mu(T_{c})$ can be different than $\mu(0)$ \cite{Chubukov16}.
To avoid this issue, one can express the superconducting transition
temperature as function of the total number of electrons in the system,
$N$, which is given by:
\begin{align}
N & =2\sum\limits _{\mathbf{k}}\left[1-T_{c}\sum\limits _{j,n}\frac{\xi_{j,\mathbf{k}}}{\omega_{n}^{2}+\xi_{j,\mathbf{k}}^{2}}\right]\nonumber \\
 & =2\pi\mathcal{V}\sum\limits _{j=1}^{2}\rho_{j,0}\left\langle \frac{1}{1+e\,^{\xi/T_{c}}}\right\rangle _{j}^{\Lambda}\text{ ,}\label{s1eq12}
\end{align}

\noindent where $\mathcal{V}$ denotes the total volume of the system
(or total area, in the 2D case). Note that, here, the upper integration
cutoff is the bandwidth $\Lambda$.

The numerical solution of Eqs. (\ref{s1eq10}) and (\ref{s1eq12})
is straightforward, and gives $T_{c}(N)$ as shown in Figs.\ref{clean}
(a) and (b) for the 2D and 3D cases, respectively. In these figures,
$T_{c}$ is normalized by $\varepsilon_{0}$ and $N$ is normalized
by the critical value $N_{c}$ at which the LT takes place, which
corresponds to $\mu(0)=\varepsilon_{0}$. For this particular figure,
we used the same density of states for both bands ($\rho_{1,0}=\rho_{2,0}$),
we set the interaction cutoff and the bandwidth to the same value
$\Omega_{0}=\Lambda=5\varepsilon_{0}$, and considered dominant intra-band
interactions $\lambda_{11}=\lambda_{22}=0.13$ with subleading inter-band
interactions, $|\lambda_{12}|\ll\lambda_{11}$. The main feature is
the enhancement of $T_{c}$ in the vicinity of the LT. Such an enhancement
is sharper for 2D bands since in this case the density of states is
discontinuous as the chemical potential crosses the band edge. 

\subsection{Asymptotic solution}

\label{sec_clean_asympt}

To set the stage for the analytic investigations of the dirty case,
here we derive an analytic asymptotic expression for $T_{c}(\mu)$
in the particular case of 2D bands. Note that, as discussed in Ref.
\cite{Trevisan_Schutt_Fernandes} and illustrated in Fig. \ref{clean},
the case of 3D bands is qualitatively similar than the 2D case. The
main quantitative differences arise from the fact that the density
of states of the 3D bands vanish smoothly at the band edge. Moreover,
because the behavior of the curves $T_{c}(\mu)$ and $T_{c}(N)$ are
very similar, as illustrated in Fig.\ref{clean}(c), we will focus
on the former.

Returning to the matrix elements $\left(\hat{A}_{\mathrm{clean}}\right)_{ij}$
in Eq.(\ref{s1eq11}), it is clear that the main effect of the LT
is on the lower integration limits $W_{i}$. Recall that $W_{1}=-\mu$
is the bottom of band $1$ and $W_{2}=-\mu+\varepsilon_{0}$ is the
bottom of band $2$. If the chemical potential was such that $\mu\gg\Omega_{0}$,
the problem would be in the high-density limit, and we would recover
the usual BCS result $\left(\hat{A}_{\mathrm{clean}}\right)_{ij}=\delta_{ij}\left(\frac{\rho_{i,F}}{\rho_{i,0}}\right)\ln\left(\frac{1.13\Omega_{0}}{T_{c}}\right)$,
where $\rho_{i,F}$ is the density of states at the Fermi level. To
capture the behavior near the LT, we first perform the energy integration
and obtain:

\begin{equation}
\left(\hat{A}_{\mathrm{clean}}\right)_{ii}=T_{c}\sum_{n}\frac{1}{\omega_{n}}\left[\text{arctan}\left(\frac{\Omega_{0}}{\omega_{n}}\right)-\text{arctan}\left(\frac{W_{i}}{\omega_{n}}\right)\right]\text{ .}\label{s1eq13}
\end{equation}

\begin{figure}[b!]
\centering \includegraphics[width=0.9\columnwidth]{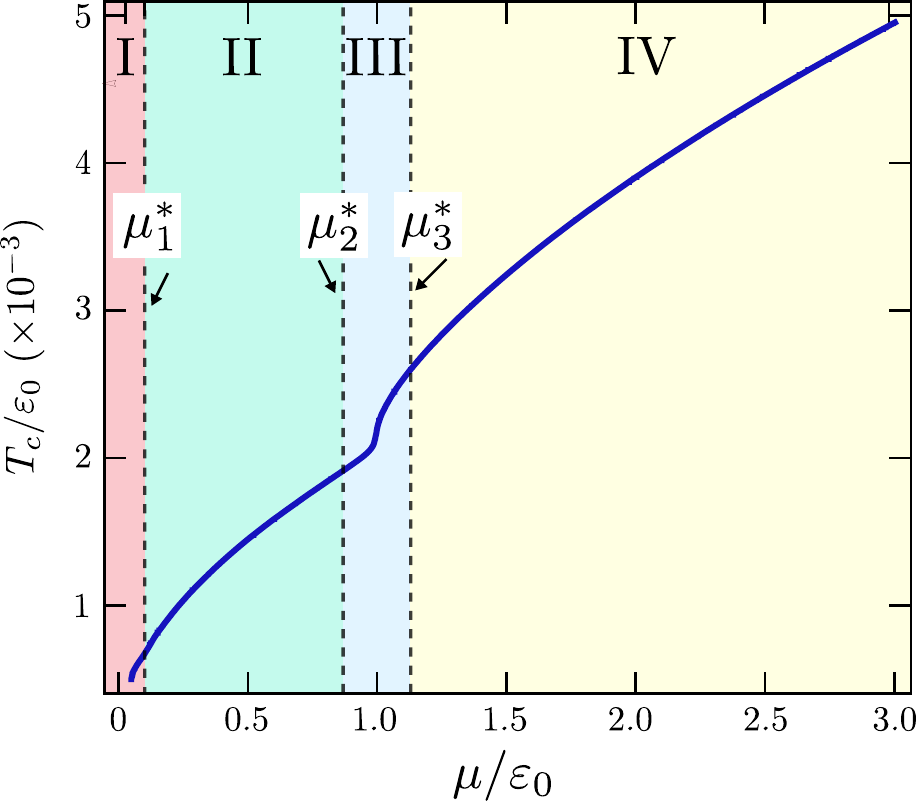}
\caption{(color online) Regions of the $(\mu,T)$ phase diagram for the calculation of the
asymptotic behavior of $T_{c}(\mu)$ in the clean and dirty regimes.
The size of the regions are exaggerated for schematic purposes. The
precise definition of each region is given in the main text.}
\label{regions} 
\end{figure}

For each band, there are two different asymptotic regimes in which
the Matsubara sum can be evaluated analytically: $\left|W_{i}\right|\ll T_{c}$
and $\left|W_{i}\right|\gg T_{c}$ (note that, in our weak-coupling
approach, $\Omega_{0}\gg T_{c}$ always). This defines $4$ regions
in the $\left(\mu,T_{c}\right)$ phase diagram, as schematically shown
in Fig. \ref{regions}:
\begin{itemize}
\item In region I, we have $-W_{1}<T_{c}$ and $W_{2}>T_{c}$. This region
corresponds to $\mu<\mu_{1}^{*}$, with $\mu_{1}^{*}\sim T_{c}\left(\mu_{1}^{*}\right)$.
\item In region II, we have $-W_{1}>T_{c}$ and $W_{2}>T_{c}$. This region
corresponds to $\mu_{1}^{*}<\mu<\mu_{2}^{*}$, with $\mu_{2}^{*}\sim\varepsilon_{0}-T_{c}\left(\mu_{2}^{*}\right)$.
\item In region III, we have $-W_{1}>T_{c}$ and $\left|W_{2}\right|<T_{c}$.
This region corresponds to $\mu_{2}^{*}<\mu<\mu_{3}^{*}$, with $\mu_{3}^{*}\sim\varepsilon_{0}+T_{c}\left(\mu_{3}^{*}\right)$.
\item In region IV, we have $-W_{1}>T_{c}$ and $-W_{2}>T_{c}$. This region
corresponds to $\mu>\mu_{3}^{*}$.
\end{itemize}
As shown in Appendix \ref{Appendix1}, we find the diagonal matrix
elements in each region: 
\begin{equation}
\left(\hat{A}_{\mathrm{clean}}\right)_{11}\sim\frac{1}{2}\begin{cases}
\ln\left(\frac{\kappa\Omega_{0}}{T_{c}}\right)+\frac{\mu}{2T_{c}}\text{,} & \hspace{-0.3cm}\text{ region I}\\[0.5cm]
\ln\left(\frac{\kappa^{2}\Omega_{0}\mu}{T_{c}^{2}}\right)\text{,} & \hspace{-0.3cm}\text{ otherwise}
\end{cases}\text{ ,}\label{s1eq14}
\end{equation}

\noindent and
\begin{equation}
\left(\hat{A}_{\mathrm{clean}}\right)_{22}\sim\frac{1}{2}\begin{cases}
\ln\left(\frac{\Omega_{0}}{\varepsilon_{0}-\mu}\right)\text{,} & \hspace{-0.3cm}\text{ regions I and II}\\[0.5cm]
\ln\left(\frac{\kappa\Omega_{0}}{T_{c}}\right)+\frac{\left(\mu-\varepsilon_{0}\right)}{2T_{c}}\text{,} & \hspace{-0.3cm}\text{ region III}\\[0.5cm]
\ln\left(\frac{\kappa^{2}\Omega_{0}(\mu-\varepsilon_{0})}{T_{c}^{2}}\right)\text{,} & \hspace{-0.3cm}\text{ region IV}
\end{cases}\hspace{-0.5cm}\text{,}\label{s1eq15}
\end{equation}

\noindent where $\kappa=2\mathrm{e}^{\gamma}/\pi\approx1.13$, with
$\gamma$ denoting Euler's constant. Solving the gap equation (\ref{gap_eq})
now corresponds to solving a simple transcendental equation, since
$\left(\hat{A}_{\mathrm{clean}}\right)_{11}$ and $\left(\hat{A}_{\mathrm{clean}}\right)_{22}$
are analytic functions of $\mu$ and $T_{c}$. This is in contrast
to the full numerical solution, which requires numerical evaluation
of Matsubara sums or energy integrations. 

\begin{figure}
\begin{centering}
\includegraphics[width=0.9\columnwidth]{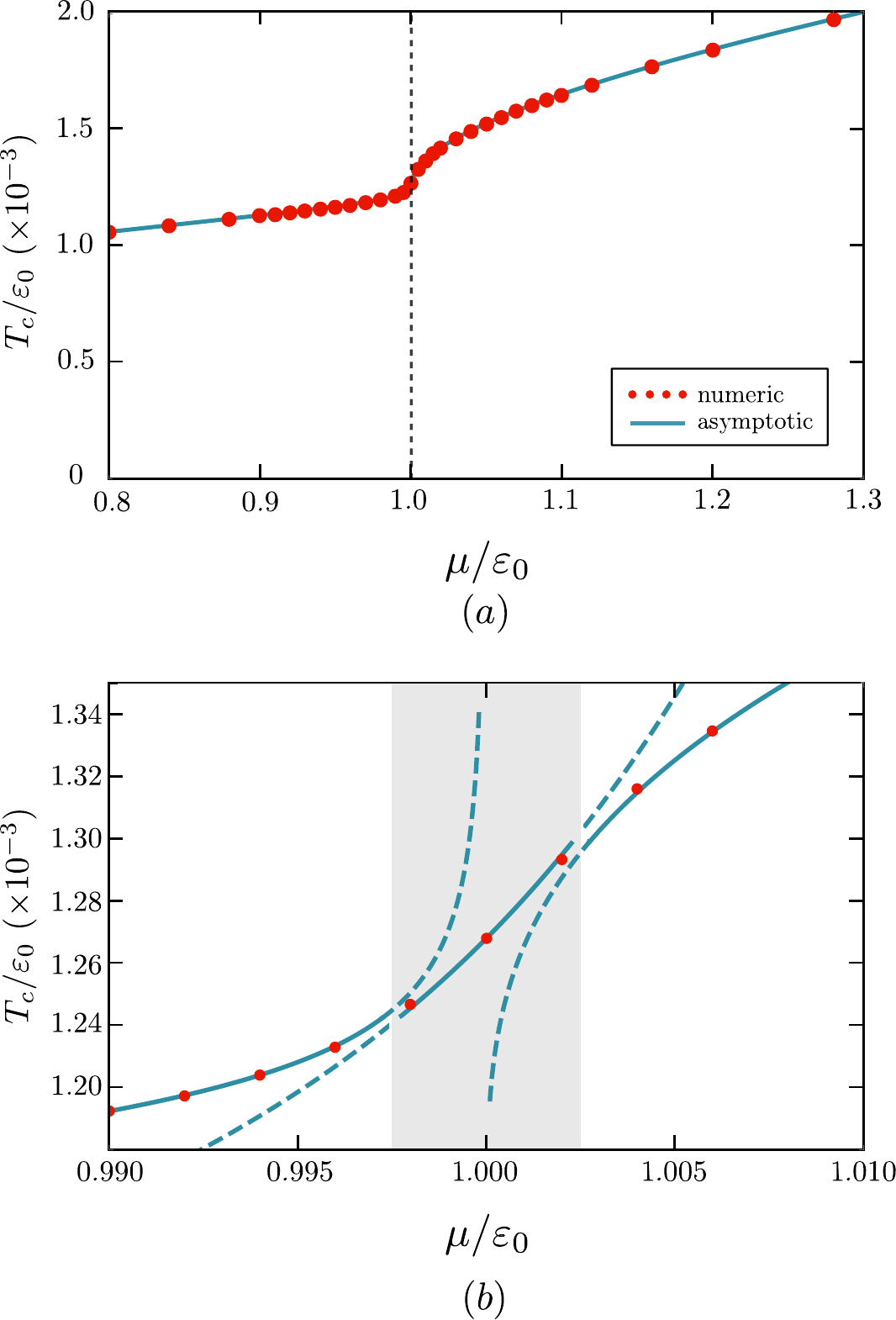}
\par\end{centering}
\caption{(color online) Comparison between the numerical (symbols) and asymptotic analytical
results (solid curve) for $T_{c}$, as function of the chemical potential
$\mu$, for the 2D clean system across the Lifshitz transition at
$\mu=\varepsilon_{0}$. Panel (b) is a zoom of panel (a) that highlights
the very narrow range of $\mu$ for which the asymptotic solutions
start to fail (gray dashed area). The parameters used here are the
same as in Fig. \ref{clean}(c). \label{fig_asymptotic_clean}}

\end{figure}

\noindent In Fig. \ref{fig_asymptotic_clean}(a), we compare the asymptotic
and numerical results for the 2D clean system, demonstrating their
excellent agreement. It is important to emphasize that, due to its
very nature, the asymptotic solution is not continuous across the
boundaries defining the different regions. In fact, as shown in Fig.
\ref{fig_asymptotic_clean}(b), some of the asymptotic solutions show
diverging behavior near the boundaries. Importantly, as highlighted
in the same figure, the ranges of $\mu$ for which the asymptotic
solutions do not behave well are very small \textendash{} in fact,
they are too small to be shown in the scale of panel (a), and are
thus omitted in that plot. Although in the clean case the advantages
of the asymptotic approach may seem rather minor, it will play an
important role in gaining insight to the behavior of the dirty system.

\section{Dirty two-band superconductor \label{sec_dirty}}

The effects of impurities in our model are captured by adding to Eq.(\ref{s1eq2})
the impurity Hamiltonian 
\begin{equation}
H_{\mathrm{imp}}=\sum\limits _{\mathbf{k},\mathbf{k'},\sigma}\sum\limits _{\alpha,\beta}W_{\alpha\beta}(\mathbf{k}-\mathbf{k'})c_{\alpha,\mathbf{k}\sigma}^{\dag}c_{\beta,\mathbf{k'}\sigma}^{\null}\text{ ,}\label{s2eq1}
\end{equation}

\noindent where $W_{\alpha\beta}(\mathbf{q})$ is the impurity potential.
Because we are interested in the case of incipient bands, we focus
on small-momentum impurity scattering. For simplicity, we consider
equal intra-band impurity potential, $v\equiv W_{11}\left(0\right)=W_{22}\left(0\right)$,
and inter-band impurity potential $u\equiv W_{12}\left(0\right)=W_{21}(0)$.

To proceed, we consider the standard self-consistent Born approximation,
as illustrated in Fig.\ref{diagrams}. The Green's function in Nambu
space is given self-consistently by Dyson's equation:

\begin{equation}
\hat{\mathcal{G}}^{-1}(\mathbf{k},\omega_{n})=\hat{\mathcal{G}}_{0}^{-1}(\mathbf{k},\omega_{n})-\hat{\Sigma}(\mathbf{k},\omega_{n})\text{ .}\label{s2eq2}
\end{equation}
where the matrix $\hat{\mathcal{G}}_{0}(\mathbf{k},\omega_{n})$ is
the Green's function of the clean system shown above in Eq.(\ref{s1eq5}),
and $\hat{\Sigma}(\mathbf{k},\omega_{n})$ is the impurity self-energy:

\begin{figure}[b!]
\centering \includegraphics[width=0.95\columnwidth]{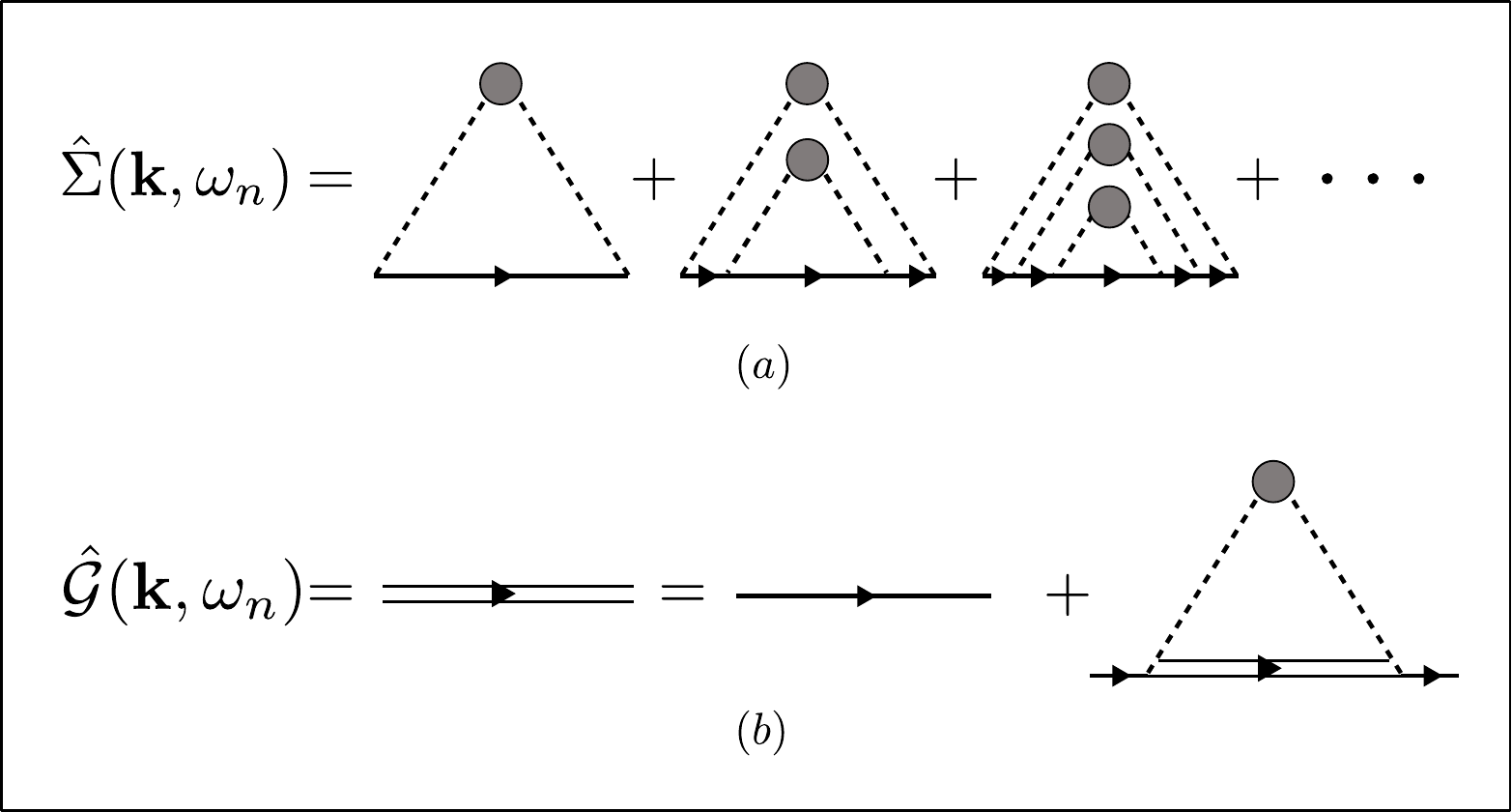}
\caption{Panel (a) shows the diagrammatic expansion for the self-energy in
the self-consistent Born approximation. Panel (b) shows the Dyson's
equation for the total Green's function according to the self-consistent
Born approximation. The solid single lines represent $\hat{\mathcal{G}}_{0}(\mathbf{k},\omega_{n})$,
while the dashed lines refer to the impurity potential $\hat{W}_{\mathbf{k},\mathbf{k'}}$.}
\label{diagrams} 
\end{figure}

\begin{equation}
\hat{\Sigma}(\mathbf{k},\omega_{n})=n_{\mathrm{imp}}\int\frac{d^{d}k'}{(2\pi)^{d}}\hat{W}_{\mathbf{k'}-\mathbf{k}}\hat{\mathcal{G}}(\mathbf{k'},\omega_{n})\hat{W}_{\mathbf{k}-\mathbf{k'}}\text{ ,}\label{s2eq3}
\end{equation}

\noindent Here, $n_{\mathrm{imp}}$ is the impurity concentration
and $\hat{W}_{\mathbf{k},\mathbf{k'}}$ represents the impurity potential
in Nambu space, 
\begin{equation}
\hat{W}_{\mathbf{k},\mathbf{k'}}=\begin{pmatrix}v & 0 & u & 0\\
0 & -v & 0 & -u\\
u & 0 & v & 0\\
0 & -u & 0 & -v
\end{pmatrix}\text{ .}\label{s2eq4}
\end{equation}

$\hat{\mathcal{G}}$ can be parametrized by the same matrix structure
as $\hat{\mathcal{G}}_{0}$ in Eq.(\ref{s1eq5}), but with renormalized
Matsubara frequencies $\tilde{\omega}_{n,j}$, energy dispersions
$\tilde{\xi}_{j,\mathbf{k}}\equiv\xi_{j,\mathbf{k}}+h_{n,j}$ and
SC gaps $\tilde{\Delta}_{j}$. As a result, we find the following
set of self-consistent equations
\begin{align}
\tilde{\omega}_{n,i} & =\omega_{n}+\sum\limits _{j}\frac{\tau_{ij}^{-1}\tilde{\omega}_{n,j}}{2}\left\langle \frac{1}{\tilde{\omega}_{n,j}^{2}+\left(\xi+h_{n,j}\right)^{2}+\tilde{\Delta}_{j}^{2}}\right\rangle _{j}^{\Lambda}\text{ ,}\label{omega_tilde}\\
\tilde{\Delta}_{i} & =\Delta_{i}+\sum\limits _{j}\frac{\tau_{ij}^{-1}\tilde{\Delta}_{j}}{2}\left\langle \frac{1}{\tilde{\omega}_{n,j}^{2}+\left(\xi+h_{n,j}\right)^{2}+\tilde{\Delta}_{j}^{2}}\right\rangle _{j}^{\Lambda}\text{ ,}\label{delta_tilde}\\
h_{n,i} & =-\sum\limits _{j}\frac{\tau_{ij}^{-1}}{2}\left\langle \frac{\xi+h_{n,j}}{\tilde{\omega}_{n,j}^{2}+\left(\xi+h_{n,j}\right)^{2}+\tilde{\Delta}_{j}^{2}}\right\rangle _{j}^{\Lambda}\text{ .}\label{h_tilde}
\end{align}

\noindent where we introduced the impurity scattering rates $\tau_{ij}^{-1}=2\pi n_{\mathrm{imp}}\rho_{j,0}\left(|v|^{2}\delta_{i,j}+|u|^{2}\delta_{\bar{i},j}\right)$,
with $\bar{i}=1(2)$ if $i=2(1)$. We also introduced here the bandwidth
$\Lambda$, which we set to be the same for both bands, for simplicity.
Since we are interested in the linearized gap equation, we can take
the limit of $\tilde{\Delta}_{j}\rightarrow0$ in the equations above.
The linear relationship between $\tilde{\Delta}_{i}$ and $\Delta_{i}$
is then given by:
\begin{equation}
\begin{pmatrix}\tilde{\Delta}_{1}\\
\tilde{\Delta}_{2}
\end{pmatrix}=\frac{1}{D_{n}}\hat{M}_{n}\begin{pmatrix}\Delta_{1}\\
\Delta_{2}
\end{pmatrix}\text{ ,}\label{s2eq8}
\end{equation}

\noindent where the matrix $\hat{M}$ is:
\begin{align}
\left(\hat{M}_{n}\right)_{ij} & =\left(1-\frac{\tau_{\bar{i}\bar{i}}^{-1}}{2}\left\langle \frac{1}{\tilde{\omega}_{n,\bar{i}}^{2}+\left(\xi+h_{n,\bar{i}}\right)^{2}}\right\rangle _{\bar{i}}^{\Lambda}\right)\delta_{i,j}\nonumber \\
 & +\frac{\tau_{ij}^{-1}}{2}\left\langle \frac{1}{\tilde{\omega}_{n,j}^{2}+\left(\xi+h_{n,j}\right)^{2}}\right\rangle _{j}^{\Lambda}\delta_{\bar{i},j}\label{s2eq9}
\end{align}

\noindent and $D_{n}\equiv\text{det}(\hat{M}_{n})$ is its determinant,
given explicitly by:

\begin{align}
D_{n} & =1-\sum\limits _{i}\frac{\tau_{ii}^{-1}}{2}\left\langle \frac{1}{\tilde{\omega}_{n,i}^{2}+\left(\xi+h_{n,i}\right)^{2}}\right\rangle _{i}^{\Lambda}+\nonumber \\
 & +\frac{\det\left(\hat{\tau}^{-1}\right)}{4}\prod_{i}\left\langle \frac{1}{\tilde{\omega}_{n,i}^{2}+\left(\xi+h_{n,i}\right)^{2}}\right\rangle _{i}^{\Lambda}
\end{align}

\noindent with $\left(\hat{\tau}^{-1}\right)_{ij}\equiv\tau_{ij}^{-1}$.
To calculate $T_{c}$, we once again relate the pair expectation value
with the anomalous Green's function, $\left\langle c_{i,-\mathbf{k}\downarrow}c_{i,\mathbf{k}\uparrow}\right\rangle =T\sum_{n}\mathcal{F}_{i}(\mathbf{k},\omega_{n})$.
Using the relationship between $\tilde{\Delta}_{i}$ and $\Delta_{i}$
above, we obtain a gap equation of the same form of Eq.(\ref{s1eq10}),
but with the matrix $\hat{A}_{\mathrm{clean}}\rightarrow\hat{A}_{\mathrm{dirty}}$.
The new matrix is given by:

\begin{equation}
\left(\hat{A}_{\mathrm{dirty}}\right)_{ij}=\pi T_{c}\sum\limits _{n}\frac{B_{i}^{(n)}}{D_{n}}\left(\delta_{ij}+C_{ij}^{(n)}\right)\label{A_dirty}
\end{equation}

\noindent where
\begin{equation}
B_{i}^{(n)}=\left\langle \frac{1}{\tilde{\omega}_{n,i}^{2}+(\xi+h_{n,i})^{2}}\right\rangle _{i}^{\Omega_{0}}\text{ ,}\label{s2eq10}
\end{equation}

\noindent and 
\begin{align}
C_{ij}^{(n)} & =-\delta_{i,j}\,\frac{\tau_{\bar{i}\bar{i}}^{-1}}{2}\left\langle \frac{1}{\tilde{\omega}_{n,\bar{i}}^{2}+\left(\xi+h_{n,\bar{i}}\right)^{2}}\right\rangle _{\bar{i}}^{\Lambda}\nonumber \\
 & +\delta_{\bar{i},j}\,\frac{\tau_{ij}^{-1}}{2}\left\langle \frac{1}{\tilde{\omega}_{n,j}^{2}+\left(\xi+h_{n,j}\right)^{2}}\right\rangle _{j}^{\Lambda}\text{ .}\label{s2eq11}
\end{align}

\begin{figure}[!t]
\centering \includegraphics[width=0.9\linewidth]{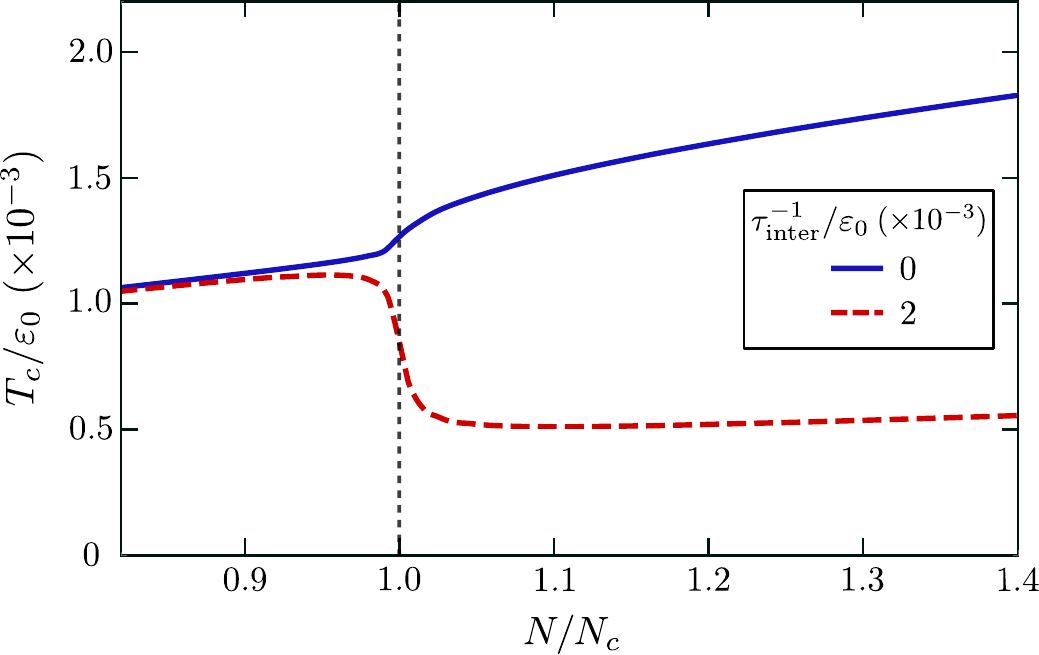}
\caption{(color online) Superconducting transition temperature $T_{c}$ as a function of the
occupation number $N$ of a dirty two-band SC with 2D bands. The dashed
line corresponds to finite inter-band impurity scattering (dirty system),
whereas the solid line corresponds to the clean system. The parameters
are the same as in Fig. \ref{clean}(c); here, the intra-band impurity
scattering is set to zero.}
\label{inter} 
\end{figure}

It is clear that, when $\hat{\tau}=0$, $\hat{A}_{\mathrm{dirty}}$
reduces to $\hat{A}_{\mathrm{clean}}$. Similarly, the equation relating
the chemical potential $\mu$ to the total number of electrons $N$
is modified to:

\begin{equation}
N=2\sum\limits _{\mathbf{k}}\left[1-T_{c}\sum\limits _{j,n}\frac{\left(\xi_{j,\mathbf{k}}+h_{n,j}\right)}{\tilde{\omega}_{n,j}^{2}+\left(\xi_{j,\mathbf{k}}+h_{n,j}\right)^{2}}\right]
\end{equation}

Solving Eqs.(\ref{omega_tilde}) and (\ref{h_tilde}) together with
the eigenvalue problem and the number equation, we can determine $T_{c}(N)$
numerically. The results, which were presented in Ref.\cite{Trevisan_Schutt_Fernandes},
reveal a pronounced suppression of $T_{c}$ at the Lifshitz transition
in the case of dominant attractive intra-band pairing and sub-dominant
repulsive inter-band pairing. In the case of sub-dominant attractive
inter-band pairing, the suppression is much milder. These results
are reproduced for completeness in Fig. \ref{inter}.

While Ref.\cite{Trevisan_Schutt_Fernandes} discussed in details the
implications of this numerical result for the understanding of the
phase diagrams of SrTiO$_{3}$ and LaAlO$_{3}$/SrTiO$_{3}$, here
we are interested in the mechanisms behind this suppression of $T_{c}$
near the Lifshitz transition, and its generalization to a wider parameter
regime that goes beyond those applicable to the materials above. To
achieve this goal, we develop now an analytical asymptotic solution
of $T_{c}$ in two different regimes: the dilute regime and the high-density
regime.

\section{Asymptotic solution of the dirty two-band superconductor \label{sec_analytics}}

Our goal here is to analytically study $T_{c}\left(\hat{\tau}^{-1}\right)$
in the different regions of the two-band superconductor $(\mu,T_{c})$
phase diagram shown in Fig. \ref{regions}. To avoid cumbersome notations,
we denote $T_{c}\left(\hat{\tau}^{-1}=0\right)\equiv T_{c,0}$, and
$\hat{A}_{\mathrm{dirty}}\equiv\hat{A}_{d}$, $\hat{A}_{\mathrm{clean}}=\hat{A}_{c}$.
Since the general function for $T_{c}\left(\hat{\tau}^{-1}\right)$
has no analytic form, we will focus here on the behavior for weak
disorder and compute $\partial T_{c}/\partial\tau_{ij}^{-1}$. This
quantity can be conveniently calculated applying Hellmann-Feynman
theorem (see for instance Refs. \cite{Rainer73,Kang16}). Recall that
$T_{c}$ is given by the solution of the linearized gap equation $\hat{\Delta}=\left(\hat{\lambda}\hat{A}_{d}\right)\hat{\Delta}$.
Let $\alpha\left(T\right)$ be the largest eigenvalue of $\left(\hat{\lambda}\hat{A}_{d}\right)$
for a given temperature $T$ and $\alpha_{0}\left(T\right)$ the largest
eigenvalue of $\left(\hat{\lambda}\hat{A}_{c}\right)$. Denote by
$\left\langle \alpha_{L}^{(0)}\right|$ and $\left|\alpha_{R}^{(0)}\right\rangle $
the left and right eigenvectors corresponding to $\alpha_{0}\left(T\right)$.
Hellmann-Feynman theorem states that
\begin{equation}
\left.\frac{\partial\alpha}{\partial\tau_{ij}^{-1}}\right|_{\tau_{ij}^{-1}=0}=\frac{\left\langle \alpha_{L}^{(0)}\left|\frac{\partial\left(\hat{\lambda}\hat{A}_{d}\right)}{\partial\tau_{ij}^{-1}}\right|\alpha_{R}^{(0)}\right\rangle }{\left\langle \alpha_{L}^{(0)}\left|\alpha_{R}^{(0)}\right.\right\rangle }\label{HF_1}
\end{equation}

Note that, because $\hat{\lambda}\hat{A}_{d}$ is generally non-symmetric,
we need to introduce both left and right eigenvectors. Recall that
we focus here in the case of fixed chemical potential $\mu$. Since
$\alpha(T_{c})=1$, using Maxwell relations, we obtain \cite{Rainer73,Kang16}:
\begin{equation}
\left.\frac{\partial T_{c}}{\partial\tau_{ij}^{-1}}\right|_{\tau_{ij}^{-1}=0}=-\frac{\left\langle \alpha_{L}^{(0)}\left|\frac{\partial\left(\hat{\lambda}\hat{A}_{d}\right)}{\partial\tau_{ij}^{-1}}\right|\alpha_{R}^{(0)}\right\rangle }{\left\langle \alpha_{L}^{(0)}\left|\alpha_{R}^{(0)}\right.\right\rangle }\frac{1}{\left.\left(\partial\alpha_{0}/\partial T\right)\right|_{T=T_{c}}}\text{ .}\label{HF2}
\end{equation}

Our goal here is to compare the changes in $T_{c}$ promoted by impurity
scattering in the high-density and dilute regimes.

\subsection{High-density regime \label{sec_high_density}}

We first discuss the high-density regime, i.e. when the system is
far from the Lifshitz transition, and the chemical potential is away
from the band edge, $\mu\gg\left\{ \Omega_{0},\varepsilon_{0}\right\} $.
This is the parameter regime most commonly studied in BCS-type approaches
to two-band superconductivity. We will recover here several results
previously published in the literature \cite{Golubov97,Efremov11,Mishra13},
but also set the stage for the analysis near the LT. 

Because $\mu\gg\left\{ \Omega_{0},\varepsilon_{0}\right\} $, the
lower cutoff of the energy integrals (\ref{integral}) is modified
according to:

\begin{equation}
\left\langle \mathcal{O}(\xi)\right\rangle _{i}^{\xi_{c}}\equiv\frac{\rho_{i,F}}{\pi\rho_{i,0}}\int\limits _{-\xi_{c}}^{\xi_{c}}d\xi\,\mathcal{O}(\xi)
\end{equation}
where $\xi_{c}$ can assume the values $\Omega_{0}$ or $\Lambda$,
and we replaced the density of states by its value at the Fermi level,
$\rho_{i,F}\equiv\rho_{i}(\xi_{F})$. In this regime, we can also
neglect the renormalization $h_{n,i}$ of the band dispersions. The
integrals that appear in the definitions of $\tilde{\omega}_{n}$
and $\hat{A}_{d}$ can then be computed in a straightforward way:

\begin{equation}
\left\langle \frac{1}{\tilde{\omega}_{n,i}^{2}+\xi^{2}}\right\rangle _{i}^{\Omega_{0}}=\left\langle \frac{1}{\tilde{\omega}_{n,i}^{2}+\xi^{2}}\right\rangle _{i}^{\Lambda}=\frac{\rho_{i,F}}{\rho_{i,0}|\tilde{\omega}_{n,i}|}
\end{equation}

As a result, the self-consistent equation for $\tilde{\omega}_{n,i}$
can be solved analytically, yielding:
\begin{equation}
\left|\tilde{\omega}_{n,i}\right|=\left|\omega_{n}\right|+\frac{1}{2}\sum\limits _{j}\tau_{ij}^{-1}\text{ ,}\label{s2eq24}
\end{equation}

In the expression above and in the remainder of this section, we renormalize
the scattering rates and coupling constants such that $\frac{\rho_{i,F}}{\rho_{i,0}}\,\tau_{ij}^{-1}\rightarrow\tau_{ij}^{-1}$
and $\frac{\rho_{i,F}}{\rho_{i,0}}\,\lambda_{ij}\rightarrow\lambda_{ij}$.
This corresponds to using the density of states at the Fermi level
$\rho_{i,F}$, instead of $\rho_{i,0}$, in the corresponding definitions,
i.e. in this section $\lambda_{ij}=-\rho_{j,F}V_{ij}$ and $\tau_{ij}^{-1}=2\pi n_{\mathrm{imp}}\rho_{j,F}\left(|v|^{2}\delta_{i,j}+|u|^{2}\delta_{\bar{i},j}\right)$.

Thus, the different components of $\left(\hat{A}_{d}\right)_{ij}=\pi T_{c}\sum\limits _{n}\frac{B_{i}^{(n)}}{D_{n}}\left(\delta_{ij}+C_{ij}^{(n)}\right)$
become:

\begin{align}
\frac{B_{i}^{(n)}}{D_{n}} & =\frac{\left(\left|\omega_{n}\right|+\frac{1}{2}\sum\limits _{j}\tau_{\bar{i}j}^{-1}\right)}{\left|\omega_{n}\right|\left(\left|\omega_{n}\right|+\frac{1}{2}\sum\limits _{j}\tau_{j\bar{j}}^{-1}\right)}\\
\frac{B_{i}^{(n)}}{D_{n}}\,C_{ij}^{(n)} & =\frac{\left(-\delta_{i,j}\tau_{\bar{i}\bar{i}}^{-1}+\delta_{\bar{i},j}\tau_{i\bar{i}}^{-1}\right)}{2\left|\omega_{n}\right|\left(\left|\omega_{n}\right|+\frac{1}{2}\sum\limits _{j}\tau_{j\bar{j}}^{-1}\right)}
\end{align}
where:

\begin{equation}
D_{n}=\frac{\left|\omega_{n}\right|\left(\left|\omega_{n}\right|+\frac{1}{2}\sum\limits _{j}\tau_{j\bar{j}}^{-1}\right)}{\prod\limits _{i}\left(\left|\omega_{n}\right|+\frac{1}{2}\sum\limits _{j}\tau_{ij}^{-1}\right)}
\end{equation}

The Matsubara sums appearing in $\hat{A}_{d}$ can be evaluated using
the result:

\begin{equation}
\sum\limits _{n}\frac{1}{|\omega_{n}|+x}\approx\frac{1}{\pi T_{c}}\left[\ln\left(\frac{\Gamma_{c}}{2\pi T_{c}}\right)-\psi\left(\frac{1}{2}+\frac{x}{2\pi T_{c}}\right)\right]\text{ ,}
\end{equation}
where $\Gamma_{c}$ is the upper cutoff of the Matsubara sum (which
is $\Gamma_{c}=\Omega_{0}\gg T_{c}$ for\textbf{ }the $B_{i}^{(n)}$
terms), and $\psi(x)$ is the digamma function. We find that $\hat{A}_{d}$
can be cast in the form:

\begin{equation}
\left(\hat{A}_{d}\right)_{ij}=\delta_{i,j}P_{i}+\delta_{\bar{i},j}Q_{i}\label{Ad_high_density}
\end{equation}
with:

\begin{align}
P_{i} & =\ln\left(\frac{\kappa\Omega_{0}}{T_{c}}\right)-\frac{\rho_{\bar{i},F}}{\rho_{1,F}+\rho_{2,F}}\left[\psi\!\left(\frac{1}{2}+\frac{\tau_{\mathrm{inter}}^{-1}}{2\pi T_{c}}\right)-\psi\!\left(\frac{1}{2}\right)\right]\\
Q_{i} & =\frac{\rho_{\bar{i},F}}{\rho_{1,F}+\rho_{2,F}}\left[\psi\!\left(\frac{1}{2}+\frac{\tau_{\mathrm{inter}}^{-1}}{2\pi T_{c}}\right)-\psi\!\left(\frac{1}{2}\right)\right]
\end{align}
where $\tau_{\mathrm{inter}}^{-1}\equiv\frac{1}{2}\left(\tau_{12}^{-1}+\tau_{21}^{-1}\right)$
is the average inter-band impurity scattering and $\kappa\approx1.13$
is the same constant that appears in Sec.\ref{sec_clean} for the
clean case.

It is clear that $\hat{A}_{d}$ depends only on the average inter-band
impurity scattering $\tau_{\mathrm{inter}}^{-1}$, i.e. $T_{c}$ is
unaffected by intra-band impurity scattering. This is not surprising,
since the gaps are isotropic and Anderson's theorem enforces that
intra-band non-magnetic impurity scattering cannot affect superconductivity.
Using these expressions, the solution of the gap equations, corresponding
to finding the largest eigenvalue of $\left(\hat{\lambda}\hat{A}_{d}\right)$,
becomes a transcendental equation that can be solved in a straightforward
way.

We now proceed to evaluate $\partial T_{c}/\partial\tau_{\mathrm{inter}}^{-1}$
using Eq. (\ref{HF2}). For convenience, we introduce the ratio between
the densities of states of the two bands to be $r\equiv\rho_{2,F}/\rho_{1,F}$.
By definition, it follows that $\lambda_{12}/\lambda_{21}=\tau_{12}^{-1}/\tau_{21}^{-1}=r$.
The largest eigenvalue of the clean gap equation is given by

\begin{equation}
\alpha_{0}=\lambda_{+}\ln\left(\frac{\kappa\Omega_{0}}{T_{c}}\right)
\end{equation}
where:

\begin{equation}
\lambda_{+}=\lambda_{0}+\sqrt{\delta\lambda^{2}+\frac{1}{r}\lambda_{12}^{2}}
\end{equation}

For simplicity of notation, here we introduced $\lambda_{0}=\frac{1}{2}\left(\lambda_{11}+\lambda_{22}\right)$
and $\delta\lambda=\frac{1}{2}\left(\lambda_{11}-\lambda_{22}\right)$.
The right and left eigenvectors are given by:

\begin{equation}
\left|\alpha_{R}^{(0)}\right\rangle =\left(\begin{array}{c}
\delta\lambda+\sqrt{\delta\lambda^{2}+\frac{1}{r}\lambda_{12}^{2}}\\
\frac{1}{r}\lambda_{12}
\end{array}\right)\label{R_eigenvector}
\end{equation}
and:

\begin{equation}
\left\langle \alpha_{L}^{(0)}\right|=\left(\begin{array}{c}
\delta\lambda+\sqrt{\delta\lambda^{2}+\frac{1}{r}\lambda_{12}^{2}}\\
\lambda_{12}
\end{array}\right)^{T}
\end{equation}

Note that the relative sign of the two components of the eigenvectors,
which correspond to the ratio between the two gaps $\Delta_{1}/\Delta_{2}$,
is determined solely by $\mathrm{sgn}\left(\lambda_{12}\right)=\mathrm{sgn}\left(\lambda_{21}\right)$,
i.e. $\mathrm{sgn}\left(\Delta_{1}/\Delta_{2}\right)=\mathrm{sgn}\left(\lambda_{12}\right)$.
As explained in Section \ref{sec_clean}, this implies that attractive
inter-band pairing interaction, $\lambda_{12}>0$, promotes a sign-preserving
$s^{++}$ state, whereas repulsive inter-band pairing interaction,
$\lambda_{12}<0$, promotes a sign-changing $s^{+-}$ state.

Next, from the definition of $\hat{A}_{d}$ in Eq. (\ref{Ad_high_density}),
we obtain:
\begin{align}
\left.\frac{\partial\left(\hat{\lambda}\hat{A}_{d}\right)}{\partial\tau_{\mathrm{inter}}^{-1}}\right|_{\tau_{\mathrm{inter}}^{-1}=0} & =\frac{1}{\left(1+r\right)}\frac{\pi}{4T_{c,0}}\times\nonumber \\
 & \left(\begin{array}{cc}
\lambda_{12}-r\lambda_{11} & -\lambda_{12}+r\lambda_{11}\\
\lambda_{22}-\lambda_{12} & -\lambda_{22}+\lambda_{12}
\end{array}\right)
\end{align}

It is straightforward to now compute $\partial T_{c}/\partial\tau_{\mathrm{inter}}^{-1}$
via Eq. (\ref{HF2}), using that $\frac{\partial\alpha_{0}}{\partial T_{c}}=-\frac{\lambda_{+}}{T_{c,0}}$.
The full expression is long and not very insightful. In the particular
case of $r=1$, however, the expression simplifies significantly and
we obtain:

\begin{equation}
\left.\frac{\partial T_{c}}{\partial\tau_{\mathrm{inter}}^{-1}}\right|_{\tau_{\mathrm{inter}}^{-1}=0}=-\frac{\pi}{8}\left[1-\frac{\mathrm{sgn}\left(\lambda_{12}\right)}{\sqrt{\left(\frac{\lambda_{11}-\lambda_{22}}{2\lambda_{12}}\right)^{2}+1}}\right]
\end{equation}

\begin{figure}
\begin{centering}
\includegraphics[width=1\columnwidth]{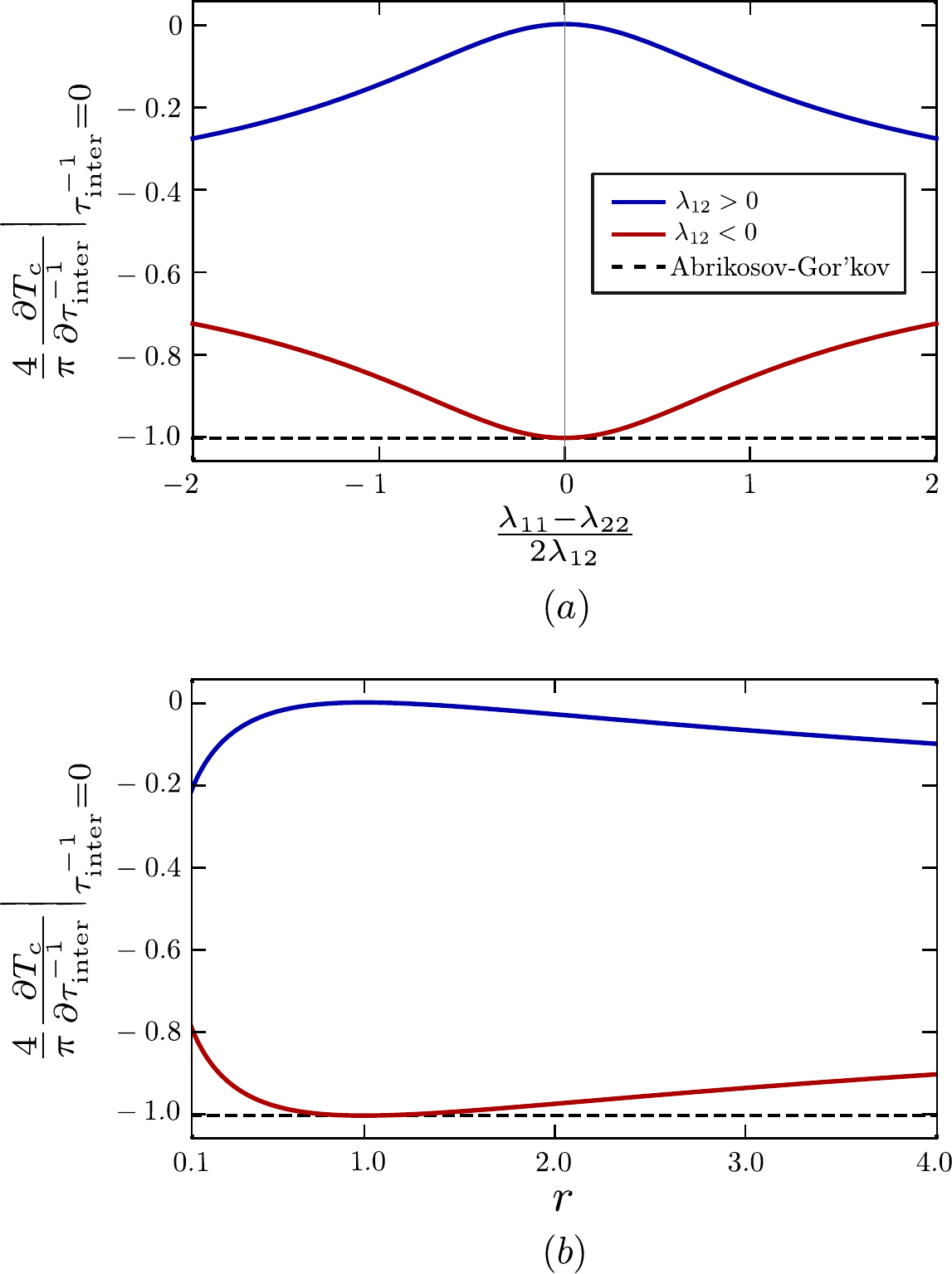}
\par\end{centering}
\caption{(color online) The rate of suppression of $T_{c}$ by inter-band non-magnetic impurity
scattering $\tau_{\mathrm{inter}}^{-1}$, $\left.\frac{\partial T_{c}}{\partial\tau_{\mathrm{inter}}^{-1}}\right|_{\tau_{\mathrm{inter}}^{-1}=0}$
for repulsive ($\lambda_{12}<0$, red curves) and attractive ($\lambda_{12}>0$,
blue curves) inter-band pairing interactions, in the high-density
regime. In panel (a), the density of states of the two bands are set
to be the same, but the intra-band pairing interactions of the two
bands, $\lambda_{11}$ and $\lambda_{22}$, are allowed to be different.
In panel (b), $\lambda_{11}$ is set to be the same as $\lambda_{22}$,
but the two density of states are allowed to be different, with $r=\rho_{2,F}/\rho_{1,F}$.
In both panels, the suppression rates are normalized by the magnitude
of the Abrikosov-Gor'kov value of $-\pi/4$ corresponding to the suppression
rate of $T_{c}$ of a single-band superconductor by magnetic impurity
scattering. \label{fig_high_density}}
\end{figure}

This expression reveals important properties of impurity scattering
in multi-band superconductors. First, as mentioned above, only inter-band
impurity scattering is pair-breaking. Second, this pair-breaking effect
takes place generically for both $s^{+-}$ and $s^{++}$ states. Indeed,
as long as $\lambda_{11}\neq\lambda_{22}$, $T_{c}$ will be suppressed
by impurities regardless of the sign of the inter-band interaction
$\lambda_{12}$. 

It is clear, however, that the suppression is stronger in the case
of repulsive interaction $\lambda_{12}<0$. Compared to the Abrikosov-Gor'kov
result for the suppression rate of $T_{c}$ by magnetic impurity scattering
in single-band $s$-wave superconductors, $\left(\frac{\partial T_{c}}{\partial\tau_{\mathrm{mag}}^{-1}}\right)_{\mathrm{AG}}=-\frac{\pi}{4}$,
it follows that $\left|\frac{\partial T_{c}}{\partial\tau_{\mathrm{inter}}^{-1}}\right|\leq\left|\frac{\partial T_{c}}{\partial\tau_{\mathrm{mag}}^{-1}}\right|$.
Note that, according to the expression for the leading eigenvector
(\ref{R_eigenvector}), the magnitudes of the two gaps are necessarily
different when $\lambda_{11}\neq\lambda_{22}$, i.e. $\left|\Delta_{1}\right|\neq\left|\Delta_{2}\right|$.
In the fine-tuned case of equal intra-band pairing interactions, $\lambda_{11}=\lambda_{22}$,
which corresponds to two gaps of same magnitudes, $\left|\Delta_{1}\right|=\left|\Delta_{2}\right|$,
$T_{c}$ for the $s^{++}$ state displays no suppression with disorder
$\tau_{\mathrm{inter}}^{-1}$, whereas $T_{c}$ for the $s^{+-}$
state displays its maximum suppression. Thus, at the same time that
$\lambda_{11}\neq\lambda_{22}$ promotes pair-breaking effects for
the $s^{++}$ state, it reduces the pair-breaking effects for the
$s^{+-}$ state. This is illustrated in Fig. \ref{fig_high_density}(a).

\begin{figure*}
\centering \includegraphics[width=0.9\linewidth]{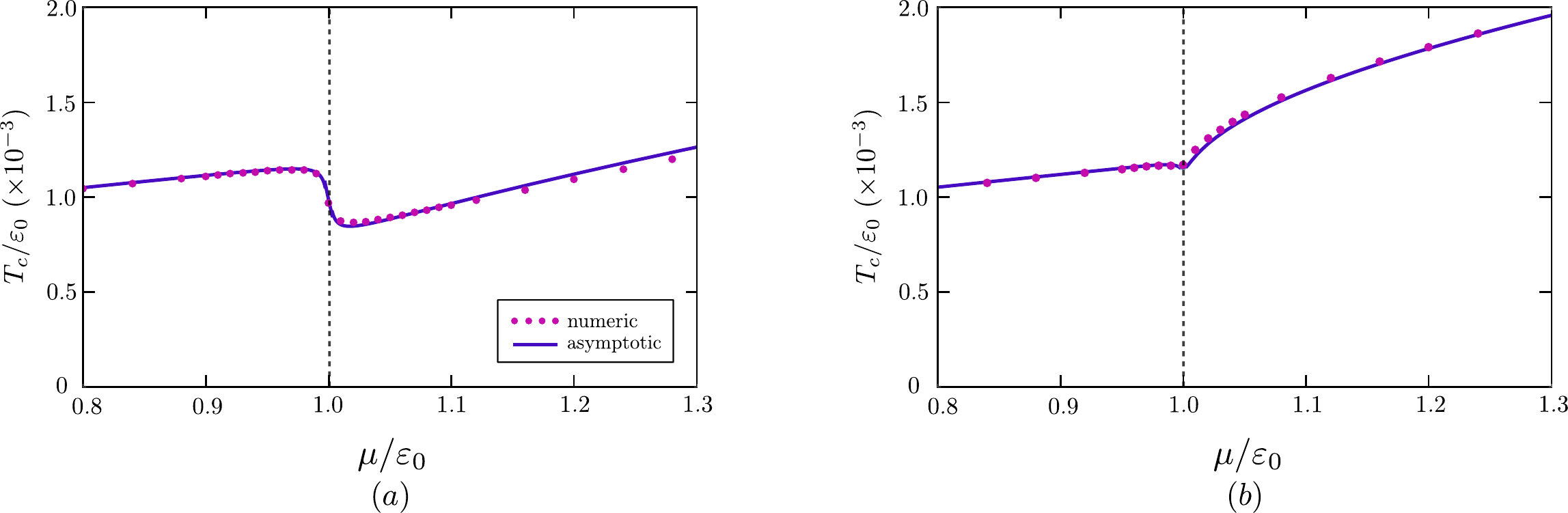}
\caption{(color online) Comparison between the numerical (symbols) and asymptotic analytical
results (solid curves) for $T_{c}$, as function of the chemical potential
$\mu$, for the 2D dirty system across the Lifshitz transition at
$\mu=\varepsilon_{0}$. The parameters are the same as in Fig. \ref{fig_asymptotic_clean}
but with $\lambda_{12}<0$ (repulsive inter-band pairing interaction)
in panel (a) and $\lambda_{12}>0$ (attractive inter-band pairing
interaction) in panel (b). The inter-band impurity scattering $\tau_{\mathrm{inter}}^{-1}$
is set to $\tau_{\mathrm{inter}}^{-1}/\varepsilon_{0}=10^{-3}$.}
\label{fig_comparison} 
\end{figure*}

The difference in the density of states between the two bands, signaled
here by $r\neq1$, plays a similar role as the difference in the intra-band
pairing interactions. For instance, if we set $\lambda_{11}=\lambda_{22}$
but consider an arbitrary $r$, we find:

\begin{equation}
\left.\frac{\partial T_{c}}{\partial\tau_{\mathrm{inter}}^{-1}}\right|_{\tau_{\mathrm{inter}}^{-1}=0}=-\frac{\pi}{8}\left[1-\frac{2\sqrt{r}\,\mathrm{sgn}\left(\lambda_{12}\right)}{1+r}\right]
\end{equation}

Once again, the suppression of $T_{c}$ for the $s^{++}$ state is
minimum (in fact, zero) when $r=1$, whereas the suppression of $T_{c}$
for the $s^{+-}$ state is maximum (and equal to the Abrikosov-Gor'kov
value) when $r=1$. This behavior is shown in Fig. \ref{fig_high_density}(b).
We emphasize that our analysis reproduces similar conclusions about
the role of impurities in multi-band superconductors that have been
previously reported elsewhere \cite{Golubov97,Efremov11,Mishra13}.

\subsection{Dilute regime \label{sec_dilute}}

Our analysis of the high-density regime reveals that impurity pair-breaking
effects on $T_{c}$ arise from the inter-band scattering rates, $\tau_{21}^{-1}$
and $\tau_{12}^{-1}$. Thus, in this subsection, to simplify the analysis,
we neglect intra-band scattering processes, and set $\tau_{11}^{-1}=\tau_{22}^{-1}=0$.
Furthermore, in the same spirit of the previous subsection, we focus
on the weak-disorder regime, in which $\tau_{12}^{-1}$ and $\tau_{21}^{-1}$
are small compared to $T_{c,0}$. Finally, we consider 2D bands, in
which case the density of states does not depend on the energy. Within
these approximations, to linear order in the scattering rates, the
renormalized Matsubara frequency in Eq.(\ref{omega_tilde}) becomes:
\begin{equation}
\tilde{\omega}_{n,i}=\omega_{n}\left(1+\frac{1}{2\pi}\tau_{i\bar{i}}^{-1}f_{n,\bar{i}}\right)\text{ ,}\label{s2eq13}
\end{equation}

\noindent where we defined the function:

\noindent 
\begin{equation}
f_{n,i}\equiv\frac{1}{\omega_{n}}\left[\text{arctan}\left(\frac{\Omega_{0}}{\omega_{n}}\right)-\text{arctan}\left(\frac{W_{i}}{\omega_{n}}\right)\right]
\end{equation}

Note that, similarly to the previous section, we neglect the renormalization
of the band due to disorder, $h_{n,i}$. As shown below, this approximation
yields very good agreement between the numerical and the asymptotic
results. Evaluating the matrix elements of $\hat{A}_{d}$ in Eq. (\ref{A_dirty}),
we find, to linear order in the impurity scattering rate:

\begin{equation}
\hat{A}_{d}=\hat{A}_{c}+\tau_{\mathrm{inter}}^{-1}\delta\hat{A}
\label{def_Ad}
\end{equation}

Here, $\hat{A}_{c}$ is the clean-case diagonal matrix discussed in
Section \ref{sec_clean_asympt}, $\tau_{\mathrm{inter}}^{-1}\equiv\frac{1}{2}\left(\tau_{12}^{-1}+\tau_{21}^{-1}\right)$
is the average inter-band impurity scattering, and $\delta\hat{A}$
is given by:

\noindent 
\begin{equation}
\left(\delta\hat{A}\right)_{ij}=\frac{1}{2\pi}\left[R_{i}\delta_{ij}+S\left(-\delta_{i,j}+\delta_{\bar{i},j}\right)\right]\label{delta_A}
\end{equation}

\noindent with 
\begin{align}
R_{i} & =-T_{c}\sum_{n}\left(\frac{\Lambda}{\Lambda^{2}+\omega_{n}^{2}}-\frac{W_{i}}{W_{i}^{2}+\omega_{n}^{2}}\right)f_{n,\bar{i}}\text{,}\nonumber \\[0.3cm]
S & =T_{c}\sum_{n}f_{n,1}f_{n,2}\text{,}\label{deltaA_aux}
\end{align}

The expressions above are obtained after two simplifications: we set
the density of states of the two bands to be equal, $\rho_{1,0}=\rho_{2,0}$,
and consider $\Omega_{0}=\Lambda$. Note that the main results presented
here do not rely on these simplifications. 

To determine analytic asymptotic expressions for the matrix elements
of $\hat{A}_{d}$, we follow the same procedure as in the clean case
as outlined in Sec.\ref{sec_clean_asympt}, and divide the $\left(\mu,T_{c}\right)$
phase diagram in four regions. The calculation is tedious but straightforward;
the resulting expressions for $R_{1}$, $R_{2}$, and $S$ are long
and shown explicitly in Appendix \ref{Appendix2}. In terms of these
expressions, finding $T_{c}$ corresponds to solving the transcendental
algebraic equation that comes from the condition that the largest eigenvalue of $\hat{\lambda}\hat{A}_{d}$ equals one (see Appendix \ref{Appendix3}).

\begin{figure*}
\begin{centering}
\includegraphics[width=1.9\columnwidth]{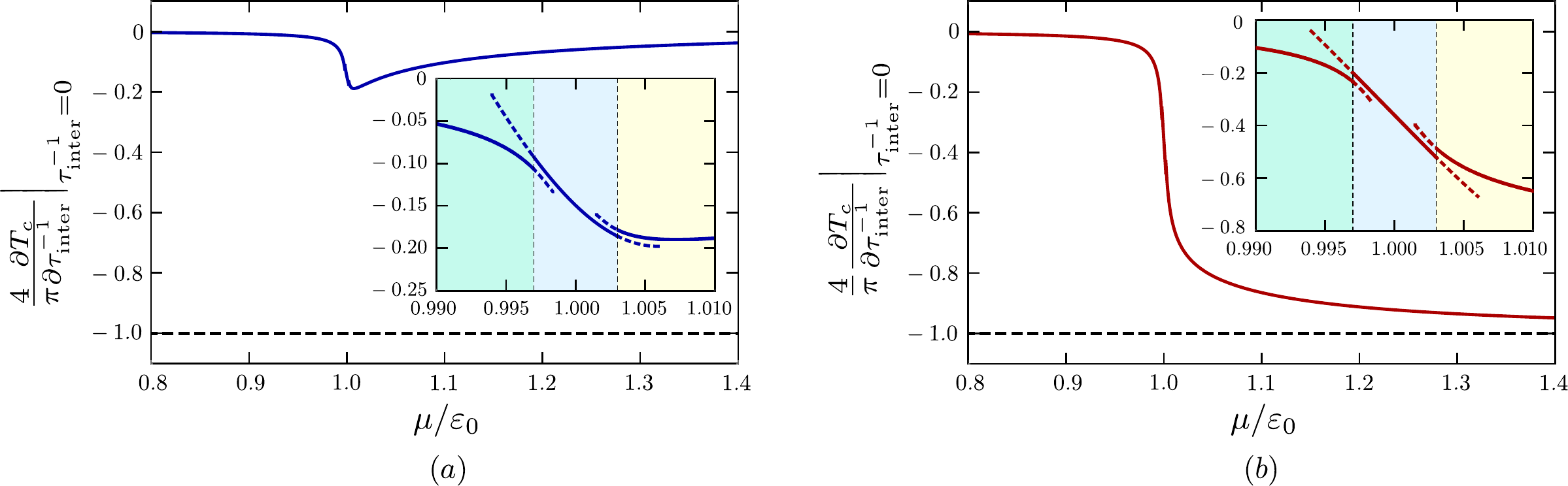}
\par\end{centering}
\caption{(color online) The rate of suppression of $T_{c}$ by inter-band impurity scattering,
$\left.\frac{\partial T_{c}}{\partial\tau_{\mathrm{inter}}^{-1}}\right|_{\tau_{\mathrm{inter}}^{-1}=0}$
for attractive ($\lambda_{12}>0$, panel (a)) and repulsive ($\lambda_{12}<0$,
panel (b)) inter-band pairing interactions, in the dilute regime.
The insets highlight the asymptotic behaviors across the boundaries
of regions II, III, and IV of Fig. \ref{regions}. In both panels,
the suppression rates are normalized by the absolute value of the
Abrikosov-Gor'kov suppression rate of $-\pi/4$, corresponding to
the case of a single-band superconductor by magnetic impurity scattering.
The parameters used here are $\rho_{1,0}=\rho_{2,0}$, $\lambda_{11}=\lambda_{22}$,
and $\lambda_{12}=\lambda_{21}$. \label{fig_low_density}}
\end{figure*}

\begin{figure}
\begin{centering}
\includegraphics[width=0.9\columnwidth]{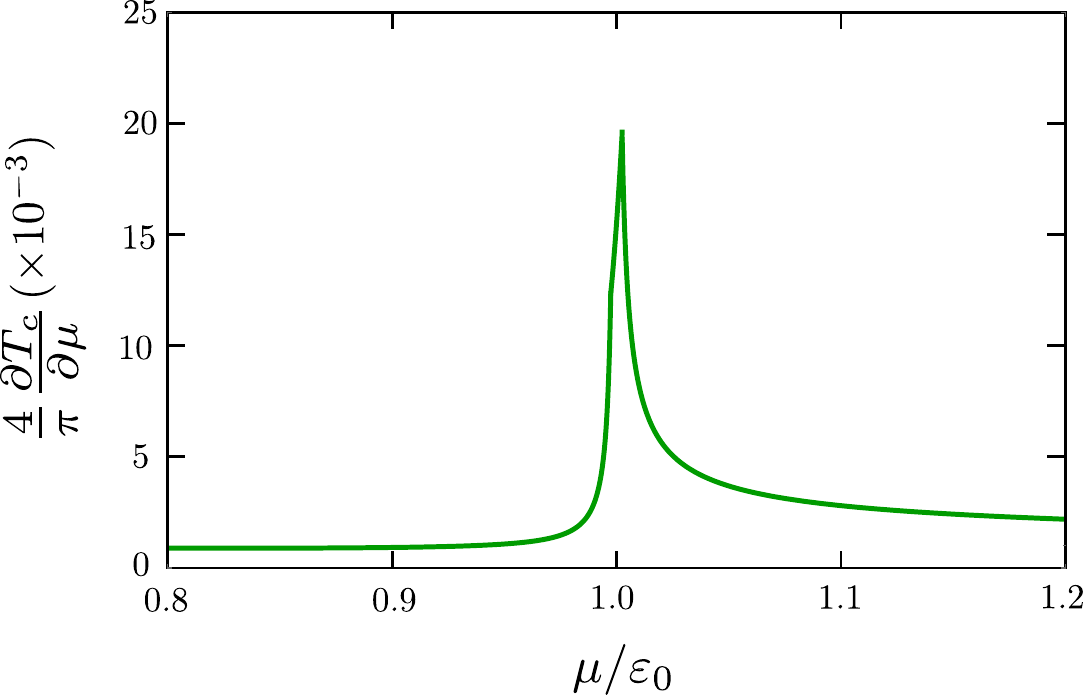}
\par\end{centering}
\caption{Rate of enhancement of $T_{c}$ by changes in the chemical potential,
$\frac{\partial T_{c}}{\partial\mu}$, for the clean 2D system. The
parameters are the same as those used in Fig. \ref{fig_low_density}.
To make the comparison with that figure more transparent, we also
normalize the rate of change of $T_{c}$ by $\pi/4$. \label{fig_dTc_dmu}}
\end{figure}

In Fig. \ref{fig_comparison}, we compare the numerical and asymptotic
analytical results for the cases of attractive and repulsive inter-band
pairing interaction. As in the clean case, the agreement between the
two methods is excellent, except in very narrow regions where the
asymptotic approximation fails. As in Fig. \ref{fig_asymptotic_clean},
these regions are too narrow compared to the scale of the plots and
are thus not shown in the plots. We note that the agreement between
the asymptotic solution and the numerical results near the LT improves
as the scattering rates becomes smaller.

In Figs. \ref{fig_low_density}(a) and (b), we plot the analytic asymptotic
behavior of $\left.\frac{\partial T_{c}}{\partial\tau_{\mathrm{inter}}^{-1}}\right|_{\tau_{\mathrm{inter}}^{-1}=0}$
as function of the chemical potential for attractive and repulsive
inter-band pairing interactions, respectively. Note that the computation of such suppression rate of $T_{c}$ from Eq.(\ref{HF2}) is straightforward and details are provided in Appendix \ref{Appendix3}. Similarly to Fig. \ref{fig_high_density},
we normalize $\partial T_{c}/\partial\tau_{\mathrm{inter}}^{-1}$
by the Abrikosov-Gor'kov suppression rate $-\pi/4$. The insets display
zooms of the behaviors of the asymptotic solutions near the LT \textendash{}
as in the analysis of previous sections, the asymptotic solutions
are not continuous across the boundaries of the different regions
of Fig. \ref{regions}. 

The results far from the LT are not surprising: before the LT, when
only one band is present, $\partial T_{c}/\partial\tau_{\mathrm{inter}}^{-1}$
is very small, since the second band is sunk below the Fermi level.
After the LT, when the second band is no longer incipient, $\partial T_{c}/\partial\tau_{\mathrm{inter}}^{-1}$
approaches the high-density values $-\pi/4$ for repulsive inter-band
interaction and $0$ for attractive inter-band pairing interaction. 

The interesting behaviors of $\partial T_{c}/\partial\tau_{\mathrm{inter}}^{-1}$
take place in the vicinity of the LT. For $\lambda_{12}<0$, we note
a very rapid increase of the magnitude of the suppression rate, despite
the fact that the second band is only incipient. On the other hand,
for $\lambda_{12}>0$, the magnitude of the suppression rate displays
a rather mild maximum when the second band crosses the Fermi level. 

The fate of the evolution of $T_{c}$ in the dirty system across the
LT depends then on the competition between two opposite effects: the
suppression of $T_{c}$ due to the pair-breaking promoted by inter-band
impurity scattering, and the enhancement of $T_{c}$ promoted by the
new electronic states that become part of the superconducting state
once the second band crosses the Fermi level. The latter effect is
illustrated in Fig. \ref{fig_dTc_dmu}, where $\partial T_{c}/\partial\mu$
obtained from the asymptotic analytical solution of the clean system
is shown. Generally, one expects that, for sufficient strong disorder,
and for a repulsive inter-band interaction, the former effect wins,
such that $T_{c}$ displays a maximum at the LT. This is indeed what
we observed in the full solution of the dirty gap equations shown
in Fig. \ref{fig_comparison}. 

\section{Concluding remarks \label{sec_concl}}

In summary, in this work we developed an asymptotic analytical framework
to investigate the behavior of the superconducting transition temperature
$T_{c}$ across a Lifshitz transition in a dirty two-band system.
Our systematic study unveiled two competing effects that influence
the evolution of $T_{c}$. The first effect arises from the fact that
the system gains energy via the opening of a superconducting gap in
the incipient band, which leads to an enhancement of $T_{c}$ (see
Fig. \ref{fig_low_density}(c)). The second effect arises because,
as soon as the second band emerges above the Fermi level and the gap
becomes non-negligible, pair-breaking effects kick in due to inter-band
impurity scattering, resulting in a suppression of $T_{c}$. While
the first effect is insensitive to the nature of pairing state \textendash{}
i.e. whether it is an $s^{++}$ state resulting from inter-band attraction
or an $s^{+-}$ state resulting from inter-band repulsion \textendash{}
the second effect is much stronger in the case of repulsive pairing
interactions. As a result, for an $s^{+-}$ superconductor with significant
impurity scattering, $T_{c}$ is expected to be maximum at the LT.
Therefore, our results offer important benchmarks to assess indirectly
from the shape of the superconducting dome whether a multi-band superconductor
is conventional (i.e. driven by attractive pairing interactions only)
or unconventional (i.e. driven by repulsive pairing interactions). 
Note that, if the impurities were magnetic, impurity scattering would be strongly pair-breaking for both attractive and repulsive inter-band interactions (see also Ref. \cite{Schimalian15}). As a result, although an explicit calculation is beyond the scope of this work, one expects a similar behavior of $T_{c}$ across the Lifshitz transition in both cases.
\begin{acknowledgments}
We thank K. Behnia, A. Chubukov, H. Faria, M. Gastiasoro, G. Lonzarich
and V. Pribiag for fruitful discussions. This work was primarily supported
by the U.S. Department of Energy through the University of Minnesota
Center for Quantum Materials, under award DE-SC-0016371 (R.M.F.).
T.V.T. acknowledges the support from the São Paulo Research Foundation
(Fapesp, Brazil) via the BEPE scholarship. 
\end{acknowledgments}

\appendix

\section{Matsubara sums for the clean case \label{Appendix1}}

Deriving an analytic expression for the matrix elements $\left(\hat{A}_{\mathrm{clean}}\right)_{ij}$
involves calculating, analytically, Matsubara sums of the type $\sum\limits _{n}\frac{1}{\omega_{n}}\text{arctan}\left(\frac{y}{\omega_{n}}\right)=\frac{\text{sign}(y)}{T_{c}}\,s_{1}\left(\frac{|y|}{T_{c}}\right)$,
where the quantity $y$ can assume the values $\Omega_{0}$, $W_{1}=-\mu$
or $W_{2}=-\mu+\varepsilon_{0}$, and 
\begin{equation}
s_{1}(|x|)\equiv2\sum\limits _{n=0}^{\infty}\frac{1}{(2n+1)\pi}\text{arctan}\left(\frac{|x|}{(2n+1)\pi}\right)\text{ .}\label{A1aeq1}
\end{equation}

We calculate an approximate expression for $s_{1}(|x|)$, taking advantage
of the asymptotic behavior of $\text{arctan}\left(\frac{|x|}{(2n+1)\pi}\right)$
in two regimes, $|x|\ll1$ and $|x|\gg1$. If $|x|\ll1$, $\frac{|x|}{(2n+1)\pi}\ll1$
for all $n$, and a Taylor expansion of $\text{arctan}\left(\frac{|x|}{(2n+1)\pi}\right)$
leads to 
\begin{equation}
s_{1}(|x|\ll1)=2\sum\limits _{l=0}^{\infty}\frac{(-1)^{l}\zeta(2l+2)\left[2^{2l+2}-1\right]}{(2l+1)(2\pi)^{2l+2}}|x|^{2l+1}\text{ ,}\label{A1aeq2}
\end{equation}

\noindent where we used the fact that 
\begin{equation}
\sum\limits _{n=0}^{\infty}\frac{1}{\left[(2n+1)\pi\right]^{k}}=\frac{\left(2^{k}-1\right)\zeta(k)}{(2\pi)^{k}}\text{ ,}\label{A1aeq3}
\end{equation}

\noindent with integer $k\geq2$ and $\zeta(k)$ denoting the Riemann
zeta function. The leading term is clearly the $l=0$:

\begin{equation}
s_{1}(|x|\ll1)\sim\frac{\left|x\right|}{4}\label{A1aeq4bar}
\end{equation}

On the other hand, if $|x|\gg1$, $\frac{|x|}{(2n+1)\pi}\gg1$ for
small values of $n$, but the ratio decreases with increasing $n$,
until it eventually behaves as $\frac{|x|}{(2n+1)\pi}\ll1$ for large
enough $n$. Denoting by $N^{*}$ the value of $n$ such that $(2N^{*}+1)\pi=\left|x\right|$,
i.e. $N^{*}=\frac{|x|}{2\pi}-\frac{1}{2}$, we approximate $\text{arctan}\left(\frac{|x|}{(2n+1)\pi}\right)$
by its Taylor expansion in powers of $1/|x|$ when $0<n<N^{*}$, and
by its Taylor expansion in powers of $|x|$ when $N^{*}+1<n<\infty$.
The result is 
\begin{align}
s_{1} & (|x|\gg1)=\sum\limits _{n=0}^{N^{*}}\frac{1}{2n+1}\nonumber \\
 & -2\sum\limits _{l=0}^{\infty}\frac{(-1)^{l}}{(2l+1)|x|^{2l+1}}\sum\limits _{n=0}^{N^{*}}\left[(2n+1)\pi\right]^{2l}\nonumber \\
 & +2\sum\limits _{l=0}^{\infty}\frac{(-1)^{l}|x|^{2l+1}}{(2l+1)}\sum\limits _{n=N^{*}+1}^{\infty}\frac{1}{\left[(2n+1)\pi\right]^{2l+2}}\text{.}\label{A1aeq4}
\end{align}

The sums over $n$ that appear in Eq.(\ref{A1aeq4}) can be evaluated
analytically:

\begin{align}
 & \sum\limits _{n=0}^{N^{*}}\frac{1}{\left[(2n+1)\pi\right]^{k}}=\nonumber \\[0.3cm]
 & \begin{cases}
\frac{(2^{k}-1)\zeta(k)}{(2\pi)^{k}}+\frac{1}{(2\pi)^{k}(|k|+1)}B_{|k|+1}\!\left(1+\frac{|x|}{2\pi}\right)\text{,} & \hspace{-0.3cm}\text{ if }k\leq0\\[0.3cm]
\frac{1}{2\pi}\left[\psi\left(1+\frac{|x|}{2\pi}\right)-\psi\left(\frac{1}{2}\right)\right]\text{,} & \hspace{-0.3cm}\text{ if }k=1\\[0.3cm]
\frac{(2^{k}-1)\zeta(k)}{(2\pi)^{k}}-\frac{1}{(k-1)!}\!\left(\frac{-1}{2\pi}\right)^{k}\psi^{(k-1)}\!\!\left(1+\frac{|x|}{2\pi}\right)\text{,} & \hspace{-0.3cm}\text{ if }k>1
\end{cases}\text{,}\label{A1aeq5}
\end{align}

\noindent and 
\begin{align}
 & \sum\limits _{n=N^{*}+1}^{\infty}\frac{1}{\left[(2n+1)\pi\right]^{k}}=\nonumber \\[0.3cm]
 & \frac{1}{(k-1)!}\!\left(\frac{-1}{2\pi}\right)^{k}\psi^{(k-1)}\!\!\left(1+\frac{|x|}{2\pi}\right)\text{ ,}\hspace{0.3cm}\text{ if }k\geq2\text{,}\label{A1aeq6}
\end{align}

\noindent where $\psi^{(k)}(x)$, $\psi(x)=\psi^{(0)}(x)$ and $B_{k}(x)$
are, respectively, the polygamma function of $k$-th order, the digamma
function, and the Bernoulli polynomials. In the limit $\left|x\right|\gg1$,
a Taylor expansion, up to order $\mathcal{O}\left(\frac{1}{\left|x\right|^k}\right)$ leads to:

\begin{align}
 & \sum\limits _{n=0}^{N^{*}}\frac{1}{\left[(2n+1)\pi\right]^{k}}\sim\nonumber \\[0.3cm]
 & \begin{cases}
\frac{1}{2\pi}\ln\left(\kappa|x|\right)\text{,} & \text{ if }k=1\\[0.3cm]
\frac{(2^{k}-1)\zeta(k)}{(2\pi)^{k}}-\frac{1}{2\pi(k-1)|x|^{k-1}}\text{,} & \text{ if }k\leq0\text{ or }k>1
\end{cases}\text{,}\label{A1aeq7}
\end{align}

\noindent and 
\begin{equation}
\sum\limits _{n=N^{*}+1}^{\infty}\frac{1}{\left[(2n+1)\pi\right]^{k}}\sim\frac{1}{2\pi(k-1)|x|^{k-1}}\text{ ,}\hspace{0.3cm}\text{ if }k\geq2\text{.}\label{A1aeq8}
\end{equation}

\noindent where we defined the constant $\kappa=2e^{\gamma}/\pi\approx1.13$,
with $\gamma$ denoting Euler's constant.

Substituting Eqs.(\ref{A1aeq7}) and (\ref{A1aeq8}) into Eq.(\ref{A1aeq4}),
we find that its second and third terms result in the same constant
$\sum\limits _{l=0}^{\infty}\frac{(-1)^{l}}{\pi(2l+1)^{2}}=\frac{\mathcal{C}}{\pi}$
($\mathcal{C}\approx0.92$ is the Catalan's constant), differing only
by a minus sign. Thus, they cancel out, and we obtain:
\begin{equation}
s_{1}(|x|\gg1)\sim\frac{1}{2}\ln\left(\kappa|x|\right)\text{ .}\label{A1aeq9}
\end{equation}

\noindent To summarize, combining Eqs.(\ref{A1aeq4bar}) and (\ref{A1aeq9}),
we have
\begin{equation}
s_{1}(|x|)\sim\begin{cases}
\frac{|x|}{4}\text{ ,} & \text{ if }|x|\ll1\\[0.3cm]
\frac{1}{2}\ln\left(\kappa|x|\right)\text{ ,} & \text{ if }|x|\gg1
\end{cases}\text{.}\label{A1aeq10}
\end{equation}

\begin{figure}[t!]
\centering \includegraphics[width=0.9\columnwidth]{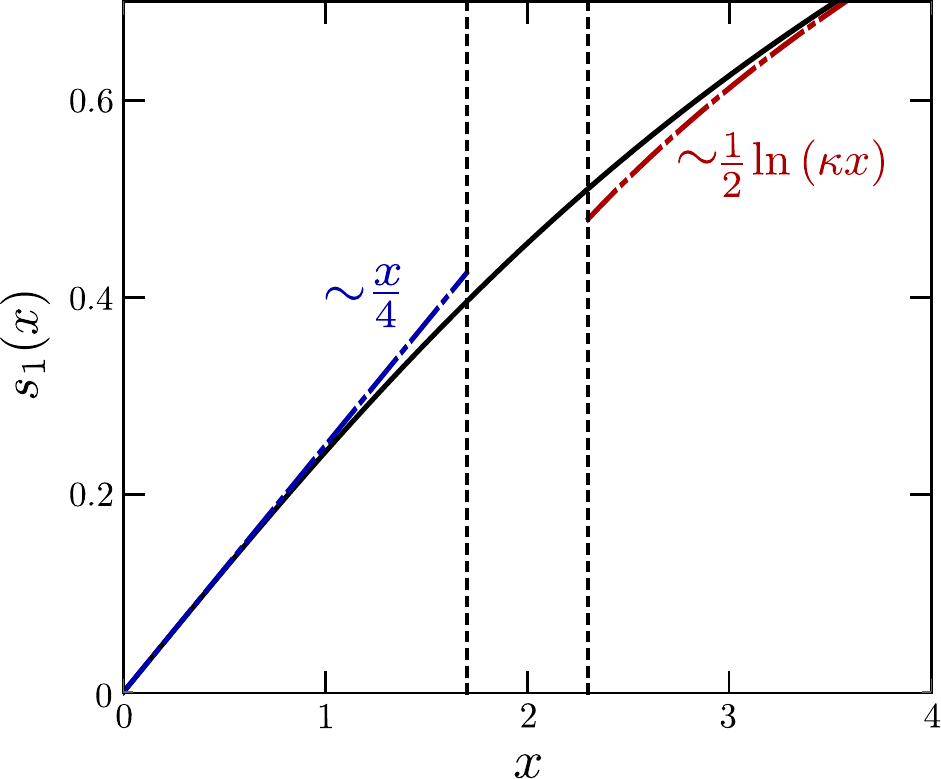}
\caption{(color online) Numerical and asymptotic solutions for the Matsubara sum (\ref{A1aeq1}).
The dot-dashed blue and red lines are the asymptotic solutions for
$|x|\ll1$ and $|x|\gg1$, while the solid line is the numerical result.
The dashed vertical lines delimit the region where the asymptotic
approximation begins to fail.}
\label{Aa_fig1} 
\end{figure}

Note that $s_{1}(|x|\rightarrow1^{+})\neq s_{1}(|x|\rightarrow1^{-})$.
This is because the asymptotic approach we described begins to fail
for $|x|$ of order one, as we can see in Fig.\ref{Aa_fig1}. As a
consequence, the asymptotic expressions for $T_{c}(\mu)$ deviate
from the numeric results when $\mu$ approaches the boundaries $\mu_{1}^{*}$,
$\mu_{2}^{*}$ and $\mu_{3}^{*}$ of the regions of the phase diagram
illustrated in Fig. \ref{regions}. At these points, either $|W_{1}|$
or $|W_{2}|$ becomes of the order of $T_{c}$.

\section{Matsubara sums for the dirty case \label{Appendix2}}

In the case of a dirty two-band SC, there are two distinct types of
Matsubara sums that we need to calculate for $\delta\hat{A}$, as
shown in Eq. (\ref{deltaA_aux}). The first are sums of the type:

\begin{align}
 & \sum\limits _{n}\frac{1}{\omega_{n}}\text{arctan}\left(\frac{y_{1}}{\omega_{n}}\right)\frac{y_{2}}{y_{2}^{2}+\omega_{n}^{2}}\nonumber \\
 & =\frac{\text{sign}(y_{1}y_{2})}{T_{c}^{2}}\,s_{2}\left(\frac{|y_{1}|}{T_{c}},\frac{|y_{2}|}{T_{c}}\right)
\end{align}
where we define:

\begin{align}
 & s_{2}(|x_{1}|,|x_{2}|)\equiv\nonumber \\
 & 2\sum\limits _{n=0}^{\infty}\frac{1}{(2n+1)\pi}\text{arctan}\!\left(\frac{|x_{1}|}{(2n+1)\pi}\right)\frac{|x_{2}|}{|x_{2}|^{2}+\left[(2n+1)\pi\right]^{2}}\text{,}\label{A1beq1}
\end{align}

The other sum is:

\begin{align*}
\sum\limits _{n}\frac{1}{\omega_{n}^{2}}\text{arctan}\left(\frac{y_{1}}{\omega_{n}}\right)\text{arctan}\left(\frac{y_{2}}{\omega_{n}}\right)\\
=\frac{\text{sign}(y_{1}y_{2})}{T_{c}^{2}}s_{3}\left(\frac{|y_{1}|}{T_{c}},\!\frac{|y_{2}|}{T_{c}}\right)
\end{align*}

\noindent where we define:
\begin{align}
 & s_{3}(|x_{1}|,|x_{2}|)\equiv\nonumber \\
 & 2\sum\limits _{n=0}^{\infty}\frac{1}{\left[(2n+1)\pi\right]^{2}}\text{arctan}\!\!\left(\!\frac{|x_{1}|}{(2n+1)\pi}\!\right)\text{arctan}\!\!\left(\!\frac{|x_{2}|}{(2n+1)\pi}\!\right)\text{.}\label{A1beq2}
\end{align}

\noindent 

\begin{figure}
\begin{centering}
\includegraphics[width=0.9\columnwidth]{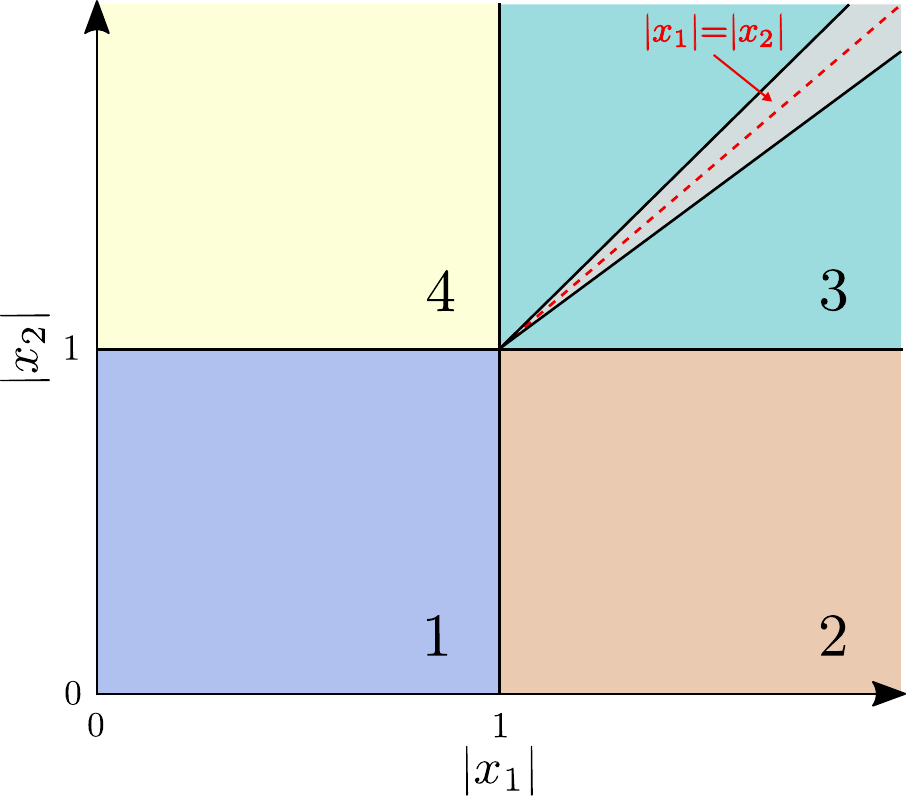}
\par\end{centering}
\caption{(color online) Different regions of the two-dimensional parameter space $|x_{1}|\times|x_{2}|$
in which the analytic expansions are performed. In region $3$, the silver area around the line $|x_{1}|=|x_{2}|$ indicates the region where the approximations lose precision, since the neglected terms of order $\mathcal{O}\left(\frac{1}{\left|x_{j}\right|}\left(\frac{\left|x_{<}\right|}{\left|x_{>}\right|}\right)^{2}\right)$, $j=1,2$, become more important. \label{fig_parameters_x1_x2}}
\end{figure}

In these expressions, both $y_{1}$ and $y_{2}$ can assume the values
$\Omega_{0}=\Lambda$, $W_{1}=-\mu$, or $W_{2}=-\mu+\varepsilon_{0}$.
To proceed with the calculation of (\ref{A1beq1}) and (\ref{A1beq2}),
we use an asymptotic approach similar to that described in Appendix
\ref{Appendix1}. In each of the four regions of the two-dimensional
parameter space $|x_{1}|\times|x_{2}|$ bounded by the lines $|x_{1}|=1$
and $|x_{2}|=1$ (see Fig.\ref{fig_parameters_x1_x2}), we substitute $\text{arctan}\left(\frac{|x_{i}|}{(2n+1)\pi}\right)$
and $\frac{|x_{i}|}{|x_{i}|^{2}+\left[(2n+1)\pi\right]^{2}}$ by their
Taylor expansions in powers of $|x_{i}|$ if $|x_{i}|\ll1$, or $1/|x_{i}|$
if $|x_{i}|\gg1$.

When $|x_{i}|\gg1$ we decompose the sums over $n$ into two contributions,
$\sum\limits _{n=0}^{\infty}f(n)=\sum\limits _{n=0}^{N_{i}^{*}}f(n)+\sum\limits _{n=N_{i}^{*}+1}^{\infty}f(n)$,
where $f(n)$ denotes any function of $n$. As in Appendix \ref{Appendix1},
$N_{i}^{*}=\frac{|x_{i}|}{2\pi}-\frac{1}{2}$ is defined such that
$(2N_{i}^{*}+1)\pi=|x_{i}|$. When both $|x_{1}|\gg1$ and $|x_{2}|\gg1$,
the decomposition is such that $\sum\limits _{n=0}^{\infty}f(n)=\sum\limits _{n=0}^{N_{<}^{*}}f(n)+\sum\limits _{n=N_{<}^{*}+1}^{N_{>}^{*}}f(n)+\sum\limits _{n=N_{>}^{*}+1}^{\infty}f(n)$,
with $N_{<}^{*}=\text{min}\{N_{1}^{*},N_{2}^{*}\}$ and $N_{>}^{*}=\text{max}\{N_{1}^{*},N_{2}^{*}\}$.
Therefore, besides the sums already calculated in Eqs. (\ref{A1aeq3}),
(\ref{A1aeq7}) and (\ref{A1aeq8}), we also need, for $\left|x_{i}\right|\gg1$,
\begin{align}
 & \sum\limits _{n=N_{1}^{*}}^{N_{2}^{*}}\frac{1}{\left[(2n+1)\pi\right]^{k}}\sim\nonumber \\[0.3cm]
 & \begin{cases}
\frac{1}{2\pi}\ln\left(\frac{|x_{2}|}{|x_{1}|}\right)\text{,} & \hspace{-0.25cm}\text{ if }k=1\\[0.3cm]
\frac{1}{2\pi(k-1)}\left[\frac{1}{|x_{1}|^{k-1}}-\frac{1}{|x_{2}|^{k-1}}\right]\text{,} & \hspace{-0.25cm}\text{ if }k\leq0\text{ or }k>1
\end{cases}\text{,}\label{A1beq3}
\end{align}

After a tedious but straightforward calculation, we then find the
following asymptotic approximations for (\ref{A1beq1}) and (\ref{A1beq2})
in each of the four asymptotic regions of the $\left(\left|x_{1}\right|,\left|x_{2}\right|\right)$
plane: \begin{widetext} 
\begin{align}
s_{2}(\left|x_{1}\right|,\left|x_{2}\right|)\approx
\begin{cases}
0\text{ ,} & \hspace{-0.25cm}\text{ if }\left|x_{1}\right|,\,\left|x_{2}\right|\ll1\\[0.3cm]
\kappa'\left|x_{2}\right|\text{ ,} & \hspace{-0.25cm}\text{ if }\left|x_{1}\right|\gg1\text{, }\left|x_{2}\right|\ll1\\[0.3cm]
\frac{1}{2\left|x_{2}\right|}\ln\left(\kappa\left|x_{<}\right|\right)-\frac{\left|x_{1}\right|}{2\left|x_{>}\right|^{2}}+\frac{\left | x_2 \right |\,\theta\left(\left|x_1\right|-\left|x_2\right| \right )}{4\left|x_{1}\right|^2}\text{ ,} & \hspace{-0.25cm}\text{ if }\left|x_{1}\right|,\,\left|x_{2}\right|\gg1\\[0.3cm]
0\text{ ,} & \hspace{-0.25cm}\text{ if }\left|x_{1}\right|\ll1\text{, }\left|x_{2}\right|\gg1
\end{cases}
\label{A1beq4}
\end{align}

\noindent and 
\begin{align}
 s_{3}(\left|x_{1}\right|,\left|x_{2}\right|)\approx
\begin{cases}
0\text{ ,} & \hspace{-0.25cm}\text{ if }\left|x_{1}\right|,\left|x_{2}\right|\ll1\\[0.3cm]
\kappa'\left|x_{2}\right|\text{ ,} & \hspace{-0.25cm}\text{ if }\left|x_{1}\right|\gg1\text{, }\left|x_{2}\right|\ll1\\[0.3cm]
\frac{\pi^{2}}{16}\!-\!\frac{(\left|x_{1}\right|+\left|x_{2}\right|)}{2\left|x_{1}\right|\left|x_{2}\right|}\!\ln\!\left(\kappa\left|x_{<}\right|\right)\!-\!\frac{1}{2\left|x_{<}\right|}+\frac{\left|x_{<}\right|}{2\left|x_{>}\right|^2}\text{ ,} & \hspace{-0.25cm}\text{ if }\left|x_{1}\right|,\left|x_{2}\right|\gg1\\[0.3cm]
\kappa'\left|x_{1}\right|\text{ ,} & \hspace{-0.25cm}\text{ if }\left|x_{1}\right|\ll1\text{, }\left|x_{2}\right|\gg1
\end{cases}
\label{A1beq5}
\end{align}
\end{widetext}

\noindent Here, we defined the constant $\kappa'=\frac{7\zeta(3)}{8\pi^{2}}\approx0.11$
and defined $|x_{<}|=\text{min}\{|x_{1}|,|x_{2}|\}$ and $|x_{>}|=\text{max}\{|x_{1}|,|x_{2}|\}$.
Recall that $\zeta(x)$ is the zeta function, $\theta\left(x\right)$ is the Heaviside step function and $\kappa\approx1.13$
is the constant defined in Appendix \ref{Appendix1}.

It is important to note that we treat the approximations we use during the derivation of Eqs.(\ref{A1beq4}) and (\ref{A1beq5}) consistently: in all the four regions of the parameter space shown in Fig.\ref{fig_parameters_x1_x2}, we kept only terms up to order $\mathcal{O}(\left|x\right|^2)$, with $\left|x\right|\ll 1$. Note that there is a small sliver region around $|x_{1}|=|x_{2}|$ in region 3 where this approximation loses precision as compared to the other regions of the ($|x_{1}|$,$|x_{2}|$) plane.
   
The matrix elements of $\delta\hat{A}$, defined in Eq. (\ref{deltaA_aux}),
are given by combinations of (\ref{A1beq4}) and (\ref{A1beq5}).
In each region of the phase diagram shown in Fig.\ref{regions},
the leading contributions yield for $R_{1}$:

\begin{widetext} 
\begin{equation}
R_{1}\sim\begin{cases}
\frac{\Omega_0+\mu-\varepsilon_0}{2\Omega_0^2}+\frac{1}{2\Omega_0}\ln\left(\frac{\varepsilon_0-\mu}{\Omega_0} \right )\text{ ,} & \text{region I}\\[0.4cm]
\frac{4\Omega_0+\mu-2\varepsilon_0}{4\Omega_0^2}+\frac{\mu\;\theta\left(\varepsilon_0-2\mu \right )}{4\left(\varepsilon_0-\mu \right )^2}-\frac{\varepsilon_0-\mu}{2W_{>}^{2}}+\frac{1}{2\Omega_0}\ln\left(\frac{\varepsilon_0-\mu}{\Omega_0} \right )-\frac{1}{2\mu}\ln\left(\frac{\mu}{W_{<}} \right )\text{ ,} & \text{region II}\\[0.4cm]
\frac{4\Omega_0-\mu}{4\Omega_0^2}-\frac{1}{2\mu}\ln\left(\frac{\kappa\mu}{T_{c}} \right )+\frac{1}{2\Omega_0}\ln\left(\frac{\kappa\Omega_0}{T_{c}} \right )\text{ ,} & \text{region III}\\[0.4cm]
\frac{4\Omega_0+\mu-2\varepsilon_0}{4\Omega_0^2}+\frac{\mu-\varepsilon_0}{2\mu^2}-\frac{1}{2\mu}\ln\left(\frac{\kappa^2\mu\left(\mu-\varepsilon_0 \right )}{T_{c}^{2}} \right )-\frac{1}{2\Omega_0}\ln\left(\frac{\kappa^2\Omega_0\left(\mu-\varepsilon_0 \right )}{T_{c}^{2}} \right )\text{ ,} & \text{region IV}\\[0.4cm]
\end{cases} \text{ .}
\end{equation}

For $R_{2}$, we find:
\begin{equation}
R_{2}\sim\begin{cases}
\frac{\varepsilon_0-\mu}{4\Omega_0^2}+\frac{1}{2\left(\varepsilon_0-\mu \right )}\ln\left(\frac{\kappa\left(\varepsilon_0-\mu \right )}{T_{c}} \right )-\frac{1}{2\Omega_0}\ln\left(\frac{\kappa\Omega_0}{T_{c}} \right )\text{ ,} & \text{region I}\\[0.4cm]
\frac{\mu+\varepsilon_0}{4\Omega_0^2}-\frac{\mu}{2W_{>}^{2}}+\frac{\left(\varepsilon_0-\mu\right)\;\theta\left(2\mu-\varepsilon_0 \right )}{4\mu^2}+\frac{1}{2\left(\varepsilon_0-\mu \right )}\ln\left(\frac{\kappa^2\left(\varepsilon_0-\mu \right )W_{<}}{T_{c}^2} \right )-\frac{1}{2\Omega_{0}}\ln\left(\frac{\kappa^2\mu\Omega_0}{T_{c}^2} \right )\text{ ,} & \text{region II}\\[0.4cm]
\frac{\Omega_0+\mu}{2\Omega_0^2}+\frac{2\kappa'\left(\varepsilon_0-\mu \right )}{T_{c}^2}-\frac{1}{2\Omega_0}\ln\left(\frac{\kappa^2\mu\Omega_0}{T_{c}^2} \right )\text{ ,} & \text{region III}\\[0.4cm]
\frac{\mu+\varepsilon_0}{4\mu^2}+\frac{\mu+\varepsilon_0+4\Omega_0}{4\Omega_0^2}-\frac{1}{2\Omega_0}\ln\left(\frac{\kappa^2\Omega_0\mu}{T_{c}^2} \right )-\frac{1}{\mu-\varepsilon_0}\ln\left(\frac{\kappa\left(\mu-\varepsilon_0 \right )}{T_{c}} \right )\text{ ,} & \text{region IV}\\[0.4cm]
\end{cases} \text{ ,}
\end{equation}

and for $S$:
\begin{equation}
S\!\sim\!\begin{cases}\!
\frac{1}{2\left(\varepsilon_0-\mu \right )}-\frac{\varepsilon_0-\mu}{2\Omega_0^2}+\frac{1}{2\left(\varepsilon_0-\mu \right )}\ln\left(\frac{\kappa\left(\varepsilon_0-\mu \right )}{T_c}\right)-\frac{1}{2\Omega_0}\ln\left(\frac{\kappa\Omega_0^2}{\left(\varepsilon_0-\mu\right)T_{c}}\right)\text{ ,} & \text{region I}\\[0.4cm]
\frac{2\mu-\varepsilon_0}{2\mu\left(\varepsilon_0-\mu \right )}\!-\!\frac{\varepsilon_0-2\mu}{2\Omega_0^2}+\!\frac{1}{2W_{<}}\!-\!\frac{W_{<}}{2W_{>}^2}\!+\frac{\varepsilon_0}{2\mu\left(\varepsilon_0-\mu \right )}\!\ln\!\left(\frac{\kappa W_{<}}{T_{c}} \right )\!-\!\frac{1}{2\mu}\!\ln\!\left(\frac{\kappa\mu}{T_{c}} \right )\!+\!\frac{1}{2\left(\varepsilon_0-\mu\right)}\!\ln\!\left(\frac{\kappa\left(\varepsilon_0-\mu\right)}{T_{c}} \right )\!-\!\frac{1}{2\Omega_0}\!\ln\!\left(\frac{\kappa^2\Omega^2\mu}{\left(\varepsilon_0-\mu \right )T_{c}^2} \right )\text{ ,} & \text{region II}\\[0.4cm]
\frac{\pi^2}{8T_{c}}-\frac{2\kappa'\left(\varepsilon_0-\mu \right )}{T_{c}^2}-\frac{1}{2\mu}+\frac{\mu}{2\Omega_0^2}-\frac{1}{2\mu}\ln\left(\frac{\kappa\mu}{T_{c}} \right )-\frac{1}{2\Omega_0}\ln\left(\frac{\kappa^3\Omega_0^2\mu}{T_{c}^3} \right )\text{ ,} & \text{region III}\\[0.4cm]
\frac{\pi^2}{4T_{c}}-\frac{1}{\mu-\varepsilon_0}-\frac{\varepsilon_0}{2\mu^2}+\frac{2\mu-\varepsilon_0}{2\Omega_0^2}-\frac{1}{2\Omega_0}\ln\left(\frac{\kappa^4\Omega_0^2\mu\left(\mu-\varepsilon_0 \right )}{T_{c}^4} \right )-\frac{1}{2\mu}\ln\left(\frac{\kappa^2\mu\left(\mu-\varepsilon_0 \right )}{T_{c}^2} \right )-\frac{1}{\mu-\varepsilon_0}\ln\left(\frac{\kappa\left(\mu-\varepsilon_0 \right )}{T_c} \right )\text{ ,} & \text{region IV}\\[0.4cm]
\end{cases} \text{ ,}
\end{equation}
\end{widetext}

\noindent where, $W_{<}\equiv\text{min}\{|W_{1}|,|W_{2}|\}$ and $W_{>}\equiv\text{max}\{|W_{1}|,|W_{2}|\}$. The order of the terms in the expressions for $R_{1}$, $R_{2}$ and $S$ are also consistent with those in Eqs.(\ref{A1beq4}) and (\ref{A1beq5}).

\section{$T_{c}(\mu)$ and $\partial T_{c}/\partial \tau_{ij}^{-1}$ in the dilute regime \label{Appendix3}}

Here, we provide more details about the calculation of the analytic asymptotic expression of $T_{c}$, as well as its suppression rate by inter-band non-magnetic impurity scattering, $\left.\frac{\partial T_c}{\partial \tau_{\mathrm{inter}}^{-1}}\right|_{\tau_{\mathrm{inter}}^{-1}=0}$, as function of the chemical potential.   

Recalling that we denote by $\alpha(T)$ the largest eigenvalue of $\hat{\lambda}\hat{A}_{d}$, where $\hat{A}_{d}$ is defined in Eq.(\ref{def_Ad}), it follows that, similarly to Sec.\ref{sec_clean_asympt}, finding $T_{c}(\mu)$ involves solving a transcendental algebraic equation $\alpha=1$, with:

\begin{equation}
\alpha=\frac{1}{2}\left[a_{11}+a_{22}+\sqrt{(a_{11}-a_{22})^{2}+4a_{12}a_{21}}\right]\text{,}\label{alpha_dirty}
\end{equation}

\noindent where we defined, in terms of the analytic expressions for $R_{i}$ and $S$ calculated in Appendix \ref{Appendix2}:
\begin{align}
a_{11} & =\lambda_{11}\left[A_{1}+\frac{\tau_{\mathrm{inter}}^{-1}}{2\pi}\left(R_{1}-S\right)\right]+\frac{\tau_{\mathrm{inter}}^{-1}}{2\pi}\lambda_{12}S\nonumber \\
a_{12} & =\lambda_{12}\left[A_{2}+\frac{\tau_{\mathrm{inter}}^{-1}}{2\pi}\left(R_{2}-S\right)\right]+\frac{\tau_{\mathrm{inter}}^{-1}}{2\pi}\lambda_{11}S\nonumber \\
a_{21} & =\lambda_{12}\left[A_{1}+\frac{\tau_{\mathrm{inter}}^{-1}}{2\pi}\left(R_{1}-S\right)\right]+\frac{\tau_{\mathrm{inter}}^{-1}}{2\pi}\lambda_{22}S\nonumber \\
a_{22} & =\lambda_{22}\left[A_{2}+\frac{\tau_{\mathrm{inter}}^{-1}}{2\pi}\left(R_{2}-S\right)\right]+\frac{\tau_{\mathrm{inter}}^{-1}}{2\pi}\lambda_{12}S \text{ ,}\label{components}
\end{align}

\noindent with $A_{1}=\left(\hat{A}_{c}\right)_{11}$ and $A_{2}=\left(\hat{A}_{c}\right)_{22}$. The resulting $T_{c}(\mu)$, for both attractive ($\lambda_{12}>0$) and repulsive ($\lambda_{12}<0$) inter-band superconducting interaction, and its comparison with the numeric solution of the gap equations are shown in Fig. \ref{fig_comparison}. 

Once we know $T_{c}(\mu)$, it is straightforward to compute $\frac{\partial T_c}{\partial \tau_{\mathrm{inter}}^{-1}}$ from Eq.(\ref{HF2}). It follows that the different terms entering Eq.(\ref{HF2}), also in terms of the analytical expressions for $R_{i}$ and $S$, are given by:

\noindent \begin{widetext} 
\begin{align}
\left\langle \alpha_{L}^{(0)}\left|\frac{\partial(\hat{\lambda}\hat{A}_{d})}{\partial\tau_{\mathrm{inter}}^{-1}}\right|\alpha_{R}^{(0)}\right\rangle =\frac{1}{2\pi} & \left\{ (1-\lambda_{11}A_{2})\left[(R_{1}-S)(\lambda_{11}-\lambda_{11}^{2}A_{2}+\lambda_{12}^{2}A_{2})+\lambda_{12}S(1+\lambda_{11}A_{1})\right]\right.\nonumber \\
 & \left.+\lambda_{12}^{2}A_{1}(R_{2}-S+\lambda_{12}SA_{2})\right\} \text{ ,}
\end{align}
and

\begin{equation}
\left.\frac{\partial\alpha_{0}}{\partial T}\right|_{T=T_{c}}=\frac{1}{2-\lambda_{11}(A_{1}+A_{2})}\sum\limits _{j=1}^{2}\left(\lambda_{11}-\lambda_{11}^{2}A_{j}+\lambda_{12}^{2}A_{j}\right)\left.\frac{\partial A_{\bar{j}}}{\partial T}\right|_{T=T_{c}}
\end{equation}
as well as

\begin{equation}
\left\langle \left.\alpha_{L}^{(0)}\right|\alpha_{R}^{(0)}\right\rangle =(1-\lambda_{11}A_{2})^{2}+\lambda_{12}^{2}A_{1}A_{2}\text{ .}
\end{equation}

\end{widetext}

\noindent In the previous equations, $A_{i}\equiv\left(\hat{A}_{c}\right)_{ii}$ and we set $\lambda_{11}=\lambda_{22}$ for simplicity. The resulting $\left.\frac{\partial T_c}{\partial \tau_{\mathrm{inter}}^{-1}}\right|_{\tau_{\mathrm{inter}}^{-1}=0}$, for both attractive and repulsive inter-band superconducting interaction, are shown in Fig. \ref{fig_low_density}.

\end{document}